\documentclass[sigplan,screen,10pt]{acmart}
\renewcommand\footnotetextcopyrightpermission[1]{}
\usepackage{enumitem}
\usepackage{multirow}
\usepackage{tcolorbox}
\usepackage{cleveref}
\usepackage{mdframed}
\usepackage{caption}
\usepackage[compact]{titlesec}
\usepackage[symbol]{footmisc}

\AtBeginDocument{%
  }

\newcommand{\smalltitle}[1]{\noindent\textbf{\textit{#1}}}

\begin{document}

\title{From Good to Great: Improving Memory Tiering Performance Through Parameter Tuning}


\author{Konstantinos Kanellis}
\authornote{Both authors contributed equally to this work.}
\email{kkanellis@cs.wisc.edu}
\affiliation{%
  \institution{University of Wisconsin-Madison}
  \city{Madison}
  \state{WI}
  \country{USA}
}

\author{Sujay Yadalam}
\authornotemark[1]
\email{sujayyadalam@cs.wisc.edu}
\affiliation{%
  \institution{University of Wisconsin-Madison}
  \city{Madison}
  \state{WI}
  \country{USA}
}

\author{Fanchao Chen}
\email{fcchen@cs.wisc.edu}
\affiliation{%
  \institution{University of Wisconsin-Madison}
  \city{Madison}
  \state{WI}
  \country{USA}
}

\author{Michael Swift}
\email{swift@cs.wisc.edu}
\affiliation{%
  \institution{University of Wisconsin-Madison}
  \city{Madison}
  \state{WI}
  \country{USA}
}

\author{Shivaram Venkataraman}
\email{shivaram@cs.wisc.edu}
\affiliation{%
  \institution{University of Wisconsin-Madison}
  \city{Madison}
  \state{WI}
  \country{USA}
}


\begin{abstract}
Memory tiering systems achieve memory scaling by adding multiple tiers of memory wherein different tiers have different access latencies and bandwidth. For maximum performance, frequently accessed (hot) data must be placed close to the host in faster tiers and infrequently accessed (cold) data can be placed in farther slower memory tiers. Existing tiering solutions employ heuristics and pre-configured thresholds to make data placement and migration decisions. Unfortunately, these systems fail to adapt to different workloads and the underlying hardware, so perform sub-optimally.

In this paper, we improve performance of memory tiering by using application behavior knowledge to set various parameters (\textit{knobs}) in existing tiering systems. To do so, we leverage Bayesian Optimization to discover the good performing configurations that capture the application behavior and the underlying hardware characteristics. We find that Bayesian Optimization is able to learn workload behaviors and set the parameter values that result in good performance. We evaluate this approach with existing tiering systems, Hemem and HMSDK. Our evaluation reveals that configuring the parameter values correctly can improve performance by 2x over the same systems with default configurations and 1.56x over state-of-the-art tiering system.

\end{abstract}

\maketitle
\pagestyle{plain}

\section{Introduction}

\textbf{Need for memory tiering.} Modern data-intensive applications like graph processing, machine learning and in-memory databases demand large amounts of memory for high performance. Due to the scaling limitation of DRAM, this demand has lead to memory becoming one of the most significant costs of datacenters~\cite{tmo2022,azurememorycosts}. One way to solve this problem is to supplement existing DRAM with memory tiers consisting of new memory technologies such as Non-Volatile Memory (NVM)~\cite{nvm2018} and CXL-based memory~\cite{cxlspec}. Memory tiering enables building systems with vast amounts of memory in a cost-effective manner.

Newer memory technologies such as NVM or CXL memory have different characteristics (speed, size, cost) compared to DRAM. The higher access latencies and lower bandwidths offered by these technologies can thus significantly degrade application performance. To alleviate this problem, there is a need to design memory tiering systems that make smart data placement decisions. They should keep frequently accessed (hot) data in faster tiers like DRAM, and infrequently accessed (cold) data in slower tiers, such as NVM or CXL.

\noindent\textbf{Limitations of existing systems.} There have been numerous works that aim to smartly place and migrate data in memory tiering systems transparently to the running applications. Unfortunately, these solutions fail to perform well under all circumstances; they either do not work well with certain applications or on certain system configurations. For example, prior work~\cite{memtis2023} has observed that HeMem~\cite{hemem2021} fails to place all the hot pages in fast tier for some workloads (also discussed in Section~\ref{sec:eval-scailp-hemem}). The are two main reasons why these systems behave sub-optimally.

\textit{First}, these solutions use heuristics, such as access frequency or recency of access, along with pre-configured thresholds to make data placement decisions. These heuristics and thresholds are optimized for the common case and not for specific workload behavior. As a result, these systems fail to correctly identify the hot pages and/or fail to migrate them in time. Memtis~\cite{memtis2023} attempts to address this limitation by dynamically adjusting the threshold for page promotion to ensure the correct set of hot pages are loaded in the fast tier. However, we find that even Memtis makes sub-optimal decisions in some cases (e.g. fails to identify write-heavy pages, see Section~\ref{sec:memtis-eval}).

\textit{Second}, the policies fail to capture and/or utilize high-level information about the application such as whether the access pattern for a region is sequential or random. For instance, HeMem migrates some write heavy pages during the initialization phase of an application which turn out to be un-useful during the more critical read phase of the application. If HeMem has knowledge about the application behavior, it could avoid migrating those pages. 

We believe that a memory tiering system that can adapt to the workload running and the underlying hardware characteristics can achieve better performance than existing solutions. 
In this paper, we try to answer the question: \textit{how much performance improvement can be achieved by taking into account application behavior while making data placement decision in memory tiering systems?} 

To this end, we test this hypothesis by tuning parameters, a.k.a \textit{knobs}, of existing memory tiering systems. Our key insight is that different data placement and migration patterns can be realized by tuning various parameters in existing tiering systems. As an example, consider HeMem~\cite{hemem2021}, which has a parameter called the \texttt{read\_hot\_threshold} which is the minimum number of accesses required to a page before considering the page a candidate for promotion from slow tier to fast tier. HeMem has another parameter called \texttt{migration\_period} which determines when the candidate pages for promotion or demotion are migrated between memory tiers. Using these two parameters, one can control which pages are considered hot, and when they are migrated. Therefore, by adjusting knob values, one could possibly realize a memory tiering system that imitates and performs close to an optimal data placement algorithm~\cite{optimal2020}. 

Manually finding the best values for the the knobs is, however, extremely challenging. Not only is it time-consuming but even experts with domain expertise could fail to identify the relationship between the application behavior, the desired memory tiering system behavior, and its parameters. So, we leverage \textit{Bayesian Optimization} to identify the best-performing tiering parameter configurations for different workloads. Bayesian optimization (BO) is effective at finding the optimal value of a function with few samples. In the past, researchers have employed BO to determine the best set of knob values in many systems such as databases~\cite{kanellis2022llamatune} and storage systems~\cite{carver2020}. Our work is the first to evaluate BO to tune knobs for memory tiering systems.

We use the optimizer to discover the best knob values for two existing tiering systems, HeMem~\cite{hemem2021} and HMSDK~\cite{hmsdk2024} under different scenarios by changing workloads, input datasets, thread counts etc. From our evaluation, we find that the optimizer is able to identify memory access patterns of applications and adjust knob values accordingly. For instance, the optimizer recognizes workloads with streaming access patterns and generates a HeMem configuration that avoids migrations of such non-beneficial pages. In comparison, HeMem with the default configuration keeps migrating pages leading to lower performance (see Table~\ref{tab:summary_performance}. Overall, we find that using an optimizer to tune tiering system parameters can yield as much as $2$x improvement. 

Finally, we analyze the best-performing configurations across scenarios using memory access patterns and migrations seen over time. By comparing these patterns with the default configuration we derive some common reasons why existing tiering systems fall short. We summarize these reasons in Section~\ref{sec:discussion} and describe some research directions which can lead to better tiering systems in the future.

\section{Background and Motivation}
\label{sec:background}

\begin{table*}[]
\scriptsize
\begin{tabular}{@{}lllllll@{}}
\toprule
   & \textbf{Access monitoring} & \textbf{Promotion heuristic}  & \textbf{Promotion threshold} & \textbf{Demotion heuristic}  & \textbf{Demotion threshold} & \textbf{Migration heuristics}  \\ \midrule
AutoTiering~\cite{autotiering2017} & Page fault    & Access recency     & Static threshold    & Access frequency   &      & Watermark (demotion)    \\ \cmidrule{1-7}
Thermostat~\cite{thermostat2017}  & Page fault     & Access frequency    & Dynamic threshold    & Access frequency  & Dynamic threshold  & Migration period  \\ \cmidrule{1-7}
AMP~\cite{adaptivepagemigration2020}  & PT scanning   & \begin{tabular}[c]{@{}l@{}}Recency/frequency/\\ random\end{tabular} & Static threshold  & \begin{tabular}[c]{@{}l@{}}Recency/frequency/\\ random\end{tabular} & Static threshold      & Migration period    \\ \cmidrule{1-7}
Multi-clock~\cite{multiclock2022} & PT scanning                                    & Recency+frequency                                                   & Static threshold                                 & Access recency                                                      & Static period                                   & \begin{tabular}[c]{@{}l@{}}Migration period (promotion)\\ Watermark (demotion)\end{tabular} \\ \cmidrule{1-7}
HeMem~\cite{hemem2021}       & PEBS                                           & Access frequency                                                    & Static threshold                                 & Access frequency                                                    & Static threshold                                & Migration period                                                                            \\ \cmidrule{1-7}
TPP~\cite{tpp2023}         & Page fault                                     & Recency                                                             & Static threshold                                 & Access recency                                                      & Static period                                   & Watermark (demotion)                                                                        \\ \cmidrule{1-7}
TMTS~\cite{tmts2023}        & PT scanning+PEBS                               & Recency+frequency                                                   & Static threshold                                 & Access recency                                                      & Static period                                   & Migration period                                                                            \\ \cmidrule{1-7}
HMSDK~\cite{hmsdk2024}       & PT scanning (DAMON)                            & Access frequency                                                    & Static threshold                                 & Access frequency                                                    & Static threshold                                & Migration period                                                                            \\ \cmidrule{1-7}
Memtis~\cite{memtis2023}      & PEBS                                           & Access frequency                                                    & Dynamic threshold                                & Access frequency                                                    & Dynamic threshold                               & Migration period                                                                            \\ \bottomrule
\end{tabular}
\vspace{1.5ex}
\caption{Summary of heuristics and thresholds used in existing tiering systems. We list only the most important parameters, however each system has a lot of more parameters such as scanning/sampling frequency, page size granularity, size of lists, etc.}
\label{tab:relatedwork}
\vspace{-10pt}
\end{table*}

\paragraph{Memory tiering.} Modern applications such as in-memory databases, web serving, graph processing and machine learning demand large memory systems for high performance. They require high bandwidth and often have low tail latency requirements. Unfortunately, scaling memory naively has become challenging because: (1) DRAM scaling has stagnated over the past few years and the cost of DRAM has been increasing making memory costs a large portion of the Total Cost of Ownership (TCO)~\cite{tmo2022}, (2) the number of channels per socket is limited by number of pins on the chip.

One way to achieve cost-effective memory scaling is through \textit{memory tiering}, which involves the organization of data across multiple \textit{tiers} of memory. New memory tiers are built using new memory technologies such as NVM~\cite{nvm2018} (lower cost/byte), CXL-based memories~\cite{cxlspec} or byte-addressable NVMe SSDs~\cite{xu2015performance}.
Unlike processor caches, in tiering, data resides in a single location (tier) in the memory hierarchy and is accessed directly from the tier it is present in. These new memory tiers exhibit higher latencies and/or lower bandwidth than locally connected DRAM~\cite{nvmperf2019,cxlperf2023}. Applications can suffer from performance degradation if they access these slower tiers frequently. Therefore, for improved performance, it is desirable that the majority of memory accesses are made to data in faster tiers. To achieve this, memory tiering systems aim to smartly place frequently accessed hot data in tiers closer to the CPU and less frequently accessed cold data in slower farther tiers.

Researchers have proposed several approaches to smartly place data in memory tiers for achieving good performance~\cite{autotiering2017,thermostat2017,nimble2019,hemem2021,tpp2023,tmts2023,memtis2023,matryoshka2024}. These systems perform two key tasks: (1) \textit{identify hot and cold data/pages}, and (2) \textit{migrate the identified hot/cold pages between tiers}. 
Ideally, for maximum performance, these systems should imitate data placement and migration that an ideal tiering algorithm such as $CH_{opt}$~\cite{optimal2020} would perform.

\paragraph{Limitations of existing tiering systems.} Existing systems, unfortunately, fall short for two main reasons: (1) they employ heuristics and static thresholds as shown in Table~\ref{tab:relatedwork} to identify and migrate hot/cold pages across tiers, and (2) they rely on low-level signals such as hardware event sampling (e.g. Intel PEBS~\cite{pebs2018}) or page table accessed (A) bit scanning (e.g. DAMON~\cite{damon2019}) or page faults which do not provide high-level information such as the access pattern of the workload.

\begin{figure}[t]
    \setlength{\abovecaptionskip}{1.5ex}
    \begin{minipage}{0.46\linewidth}
        \centering
        \includegraphics[width=\linewidth]{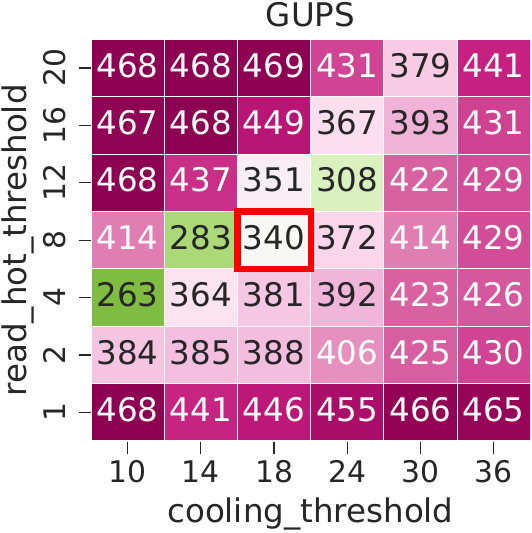}
    \end{minipage}
    \hspace{1em}
    \begin{minipage}{0.46\linewidth}
        \centering
        \includegraphics[width=\linewidth]{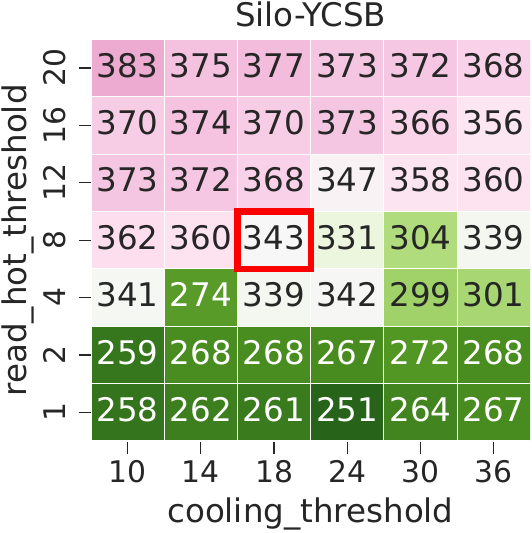}
    \end{minipage}
    \caption{Execution time (in seconds) of \textit{GUPS} (left) and \textit{Silo} (right) workloads, when we tweak two HeMem parameters. Default configuration execution time is shown in red box.}
    \label{fig:grid-search}
\end{figure}

\paragraph{Case study.} To understand the extent to which heuristics and especially static thresholds could affect memory tiering performance, we perform a simple case study using HeMem~\cite{hemem2021} on an NVM machine. HeMem and other tiering systems often include several parameters, a.k.a \textit{knobs}, such as hotness thresholds and migration period which control the system behavior (see ~\cref{tab:hemem_knobs}). Developers of these systems perform limited sensitivity studies to find the best values for these parameters which we refer to as the default values. We consider just two of HeMem's configuration parameters: \texttt{read\_hot\_threshold}, which is used to determine when a page is hot (default is $8$ sampled accesses), and \texttt{cooling\_threshold}, which controls when page counts are cooled, i.e., when page counts are halved so as to imitate an exponential moving average (default is $18$ sampled accesses).
We perform a straightforward grid search, where we tweak the values of these knobs to explore how larger and smaller values affect the execution time of a given workload; we use the default value for the remaining knobs.
Figure~\ref{fig:grid-search} shows the execution time for GUPS~\cite{plimpton2006simple} and Silo~\cite{tu2013speedy} (running a read-only YCSB-C workload) under different HeMem parameter configurations.
We highlight the default HeMem configuration with a red box.

From our results, we make the following observations.
First, we observe that different HeMem configurations result in large variations in workload performance (i.e., elapsed time).
Second, we observe that there exist configurations that perform much better compared to the default HeMem configuration for a given workload.
For example, the best-performing configuration delivers up to $29\%$ performance improvement compared to the default for GUPS and the improvement is $36\%$ for Silo.

\vspace{-1ex}
\begin{mdframed}[linecolor=black,linewidth=1pt,innerleftmargin=8pt,innerrightmargin=8pt,innertopmargin=4pt,innerbottommargin=4pt, skipabove=8pt]
\textbf{Insight \#1} \textit{Memory tiering system parameters such as thresholds and migration periods are critical for performance. Setting these parameters to the appropriate values can yield significant performance gains.}
\end{mdframed}
\vspace{1.5ex}

Third, we observe that the knob values that improve performance are quite different across our workloads, GUPS and Silo.
For GUPS, better performance is achieved with a large read hot threshold ($12-20$), while for Silo performance is improved when using small read hot threshold ($1-2$).
Existing solutions such as HeMem have a myopic view wherein they fail to capture the high-level workload behavior and the characteristics of the underlying hardware. Thus, these systems fail to adapt to the environment they are running in and behave the same way under all scenarios which leads to sub-optimal decisions.\\

\vspace{-1.5ex}
\begin{mdframed}[linecolor=black,linewidth=1pt,innerleftmargin=8pt,innerrightmargin=8pt,innertopmargin=4pt,innerbottommargin=4pt, skipabove=8pt]
\textbf{Insight \#2} \textit{The best values of these parameters depend on the workload. Since these parameters control the data placement and migration decisions made by the memory tiering system, they can be modified to express the desired migration sequence/pattern for an application.}
\end{mdframed}
\vspace{1.5ex}

Memtis~\cite{memtis2023} makes a similar observation that static thresholds do not reflect the workload behavior and proposes a strategy to dynamically decides the thresholds for hot, warm and cold pages based on the workload access distribution. This allows Memtis to outperform previously proposed tiering systems. However, we find that dynamically adjusting only the hot/warm/cold thresholds is insufficient in many cases. Memtis also fails to consider the characteristics of the underlying system. The cost of page migrations can vary depending on the available bandwidth and volume of competing traffic. Since Memtis does not consider this cost while making decisions, we see that it ends up making disadvantageous migrations. Also, Memtis uses static values for various other parameters such as the \texttt{cooling\_threshold} and \texttt{migration\_period} which leads to poor performance under many circumstances. (more details in Section~\ref{sec:memtis-eval})

Based on these observations we see two key challenges for memory tiering systems: first, it is important to efficiently \emph{discover good performing parameter configurations} for existing tiering systems that can capture the application behavior and the underlying hardware characteristics. Second, it is crucial to \emph{analyze why certain parameter configurations perform well}. Analyzing the best performing configuration can help us understand workload-specific migration patterns which lead to good performance and yield guidelines which can influence the design of future memory tiering systems.

\section{Tuning Tiering System Parameters}
\label{sec:proposal}

\begin{table*}
\footnotesize
\centering\centering
\begin{tabular}{lllll}
\toprule
\textbf{Knob Name}  & \textbf{Default} & \textbf{Min}  & \textbf{Max} & \textbf{Description} \\
\midrule
sampling\_period & 5000 & 100  & 10000 & Number of memory load events to trigger sampling \\
write\_sampling\_period & 10000   & 1000 & 20000 & Number of store instructions to trigger sampling \\
read\_hot\_threshold & 8  & 1   & 30   & Minimum number of read access samples per page to classify it hot \\
write\_hot\_threshold & 4  & 1   & 30 & Minimum number of write samples per page to classify it hot \\
cooling\_threshold & 18   & 4    & 40   & Number of sampled accesses to trigger page access count cooling\\
migration\_period & 10   & 10    & 5000    & Interval of migration thread executions (ms) \\
max\_migration\_rate & 10       & 2    & 20    & Maximum migration rate allowed (GiB/s) \\
cooling\_pages  & 8192  &   1024   &   65536  & Number of pages cooled at a time \\
hot\_ring\_reqs\_threshold & 1024    & 128    & 4096    & Number of hot pages processed at a time  \\
cold\_ring\_reqs\_threshold & 32    & 8    & 256    &  Number of cold pages processed at a time \\
\bottomrule
\end{tabular}
\vspace{1ex}
\caption{HeMem parameter knobs, their defaults, and ranges input to the optimizer}
\label{tab:hemem_knobs}
\vspace{-3ex}
\end{table*}

As discussed in the previous section, finding the right set of values for tiering system parameters can result in major performance improvements.
Prior works~\cite{hemem2021,tmts2023} use domain expertise and sensitivity studies to find the "best" parameter values but this does not scale well.
Additionally, manually finding the best values for every workload and hardware platform can be extremely challenging.

Some tiering systems also have a large parameter space (e.g., many parameters with large value ranges), with several parameters being correlated with each other~\cite{hemem2021, hmsdk2024}.
Therefore it is possible that only a specific combination of right parameter values may actually bring performance benefit, meaning that parameters should not be tuned independently from each other.
Moreover, there might exist \textit{hidden} parameters, i.e., parameters that are not explicitly exposed by the tiering engine.
Such parameters are often neglected, but if considered, may actually bring meaningful improvements (as we show for Silo in Section~\ref{sec:eval-scailp-hemem}).
Note that the above challenges are not limited to memory tiering engines, but have also been observed in many other systems, including DBMSs~\cite{kanellis2022llamatune, tuna2025} and big-data analytics frameworks, like Spark~\cite{tuneful2020}, or key-value stores, like RocksDB~\cite{dremel22, kv-rl-tuning23}.

Furthermore, understanding the relationship between application behavior, parameter values and their effect on the migration decisions can be quite tricky.
To address this, we need a tuning process that is more principled compared to traditional grid- or random- search methods, which explore the parameter space in a rather unguided fashion.
Instead, we need a process that (1) \textit{maximizes} the information gained from each configuration evaluation, (2) is very \textit{sample-efficient} and (3) continuously tries to \textit{learn} how the tiering engine behaves.

\subsection{Proposal: Leverage Bayesian Optimization}
\label{sec:bo}

To this end, we propose employing Bayesian Optimization (BO).
BO is a sequential, model-based optimization process that aims to find the (global) optimum of a black-box function using as few trials as possible~\cite{snoek2012practical}.
It is typically used when the black-box function is expensive to evaluate.
Notably, BO does not make any a-priori assumptions regarding the shape/behavior of the function (e.g., convex, monotonic, etc), making it a prime candidate to use in our case~\cite{bo-survey}.

In our work, BO can be used to identify the best-performing configuration (global optimum) for our desired tiering engine (black-box function).
To find a best-performing configuration, we need to determine an appropriate value for each parameter based on the workload and hardware.
Formally, given a set of tiering engine parameters $\theta_1, ..., \theta_n$ along with their domains (i.e., range of values), $\Theta_1, ..., \Theta_n$, the tiering system parameter space can be defined as $\mathbf{\Theta} = \Theta_1 \times ... \times \Theta_n$.
We want to find a configuration $\mathbf{\theta^*} \in \mathbf{\Theta}$ that minimizes the execution time of the workload $f$:
\begin{equation*}
    \mathbf{\theta^*} = \arg \min_{\theta \in \mathbf{\Theta}}f(\theta)
\end{equation*}

A Bayesian optimizer consists of two core components: the surrogate model and the acquisition function.
The surrogate model is a machine-learning model that models the behavior of the system (i.e., tiering system), and can predict its performance given a specific configuration.
Initially the model has no knowledge of the system, but with more and more observations (i.e., evaluated configurations), it quickly "learns" how the tiering engine behaves.
The acquisition function is used by the optimizer to decide which configuration is more promising to evaluate next. This is done by balancing exploration (i.e., evaluate a configuration on a previously unexplored region of the parameter space) with exploitation (i.e., try to improve an already good-performing configuration by making small tweaks on parameter values).

\smalltitle{Tuning Pipeline.} Given a fixed workload, we first launch a tuning session, where we (1) set our tiering sytem to a configuration (initially default), (2) execute the workload, and (3) measure its execution time.
Then, we feed the evaluated configuration alongside the execution time back to the optimizer.
The optimizer updates the surrogate model with the new observation and selects the best candidate configuration to be evaluated next (via the acquisition function).
We repeat the above process, always using the new configuration suggested by the optimizer, until we exhaust the tuning budget (e.g., number of iterations).
To balance between exploration and exploitation, we also (1) initially bootstrap the optimizer with some randomly-sampled configurations, and (2) force the optimizer to periodically suggest a random configuration (i.e., random config probability).

\smalltitle{BO Algorithm.} In this work we use the state-of-the-art Sequential Model-based Algorithm Configuration (SMAC) framework~\cite{hutter2011sequential}, which is a popular BO-based optimizer~\cite{bo-survey}.
SMAC utilizes a Random Forest (RF) as the surrogate to model the parameter space; this enables it to handle large high-dimensional spaces efficiently.
As shown by prior work~\cite{kanellis2022llamatune}, when used to tune complex systems, SMAC manages to find best-performing configurations within few evaluations, even when having hundreds of knobs, with a subset of them correlated with each other.
Considering our previous case study in Figure~\ref{fig:grid-search}, SMAC was able to find the best-performing configuration for GUPS within $10-16$ iterations, making it $2.5-4\times$ more sample-efficient.

An additional benefit of using SMAC's random forest surrogate model is its inherent ability to determine which tiering system knob(s) are more important.
This is performed by computing an \textit{importance score} for every knob, based on its impact to the workload execution time (similar to~\cite{carver2020, kanellis2020too}).
In particular, for each knob $k$, we fix the values of other knobs at their default value and then randomly sample across values of $k$ to observe the impact on workload performance (via surrogate model prediction).
Assuming enough observations, the surrogate model can make accurate predictions.
While one should not blindly rely on these importance scores, these can be helpful when trying to understand why some configurations perform better than others, like in Section~\ref{sec:eval-scailp-hemem}.


\subsection{Tiering engine under the spotlight: HeMem}

We use the proposed optimization pipeline to tune knobs of two tiering engines, HeMem~\cite{hemem2021} and HMSDK~\cite{hmsdk2024}. In this section, we give a brief overview of HeMem which will be helpful to understand our analysis in Section~\ref{sec:evaluation}. We discuss HMSDK later in Section~\ref{sec:hmsdk_tuning}.
HeMem is a user-space memory manager for tiered memory systems that is dynamically and transparently linked into applications. HeMem intercepts \texttt{mmap()}, handling page allocations pertaining to anonymous memory. 
Table~\ref{tab:hemem_knobs} lists all configuration knobs we consider, which are also discussed in the following paragraphs.

HeMem monitors page accesses using hardware event sampling such as Intel PEBS~\cite{pebs2018}. HeMem samples L3 load misses as well as all store instructions. Based on the samples gathered, HeMem maintains an access count per page. HeMem uses separate counters for reads and writes. If the page access count exceeds a pre-determined threshold, it classifies the page as hot; else considers the page cold. It uses different thresholds for reads (\texttt{read\_hot\_threshold}) and writes (\texttt{write\_hot\_threshold}).

HeMem uses a page count cooling mechanism to give importance to latest accesses. When the access count of any page reaches a \texttt{cooling\_threshold}, page cooling is triggered which halves access count of all pages. If the access count falls below the hot threshold after cooling, the page is then considered cold. To avoid scanning all pages, HeMem cools pages in batches whose size is determined by another parameter called \texttt{cooling\_pages}. This is one of the (hidden) parameters that is not discussed in the paper but is part of the implementation.

HeMem includes a background migration thread that executes periodically every \texttt{migration\_period}. The migration thread promotes hot pages on slower tiers and demotes cold pages from faster tiers. HeMem marks pages as write-protected when they are being migrated to avoid races between application writes and the data movement. Applications could therefore stall waiting for pages to be migrated.

\smalltitle{Deployment Issues.}
During our initial tests with HeMem, we observed some abnormal behavior, which often led to workload crashes or poor performance, affecting our results.
In an effort to achieve a more accurate evaluation, we tried to address these issues without compromising the original HeMem implementation design and functionality.
Below, we briefly summarize these issues and our corresponding fixes:

\begin{enumerate}[leftmargin=1.2em]
    \item \textit{High sampling overhead}: In the original code, PEBS hardware generated an interrupt for every sample, incurring high overhead.
    This is due to certain parameters requested from PEBS (\texttt{PERF\_SAMPLE\_WEIGHT}).
    We modified this parameter, so that the hardware generates much fewer interrupts without affecting sampling accuracy.

    \item \textit{Migrate-free page race condition}: HeMem's original implementation does not properly protect against cases where a page is freed by the workload, while HeMem tries to migrate at the same time. This causes some workloads to completely crash. We fixed this issue by introducing lightweight synchronization. 

    \item \textit{Large minimum allocation size}: By default, HeMem is configured to only handle page allocation greater than $1$GiB in size. We changed this to $128$MiB, as we observed that some of our workloads do not make allocations that large, making HeMem redundant for those cases.

    \item \textit{Read/write sampling frequency separation}: We introduce a new parameter called \texttt{WRITE\_SAMPLING\_PERIOD} that controls PEBS sampling frequency of store instructions. This allows HeMem to exert different importance for read and write operations.
\end{enumerate}

\section{Evaluation \& Analysis}
\label{sec:evaluation}

Through our evaluation and analysis, we try to answer the following questions:
\begin{enumerate}[leftmargin=1.2em]
    \item How much performance improvement can the best-performing configuration achieve for HeMem across workloads?
    \item What are the important differences between the default and the best-performing knob values? How does this translate into additional performance?
    \item Is there a single best-performing configuration that would work well in all scenarios?
    \item Under what circumstances do tiering systems leave unexploited performance on the table?
    \item How do the above results and observations generalize to a different tiering system?
    \item How does the performance of a tuned tiering system compare against state-of-the-art designs like Memtis which dynamically adjust certain knobs (e.g. \texttt{hot\_threshold})?
\end{enumerate}

\begin{table}[t]
\footnotesize
\centering
\begin{tabular}{@{}lccc@{}}
\toprule
\textbf{Specification} & \textbf{pmem-large} & \textbf{pmem-small} & \textbf{NUMA} \\ \midrule
Number of cores                   & 24      & 16     & 20     \\
Processor generation               & Icelake & Cascadelake & Skylake \\
Processor frequency (GHz)          & 3       & 3.2    & 2.2    \\
L3 cache size (MB)                 & 18      & 11     & 13.75  \\
\hline
Far memory type                    & Optane  & Optane & NUMA   \\
Max near memory size (GB)          & 96      & 32     & 96     \\
Max far memory size (GB)           & 128     & 128    & 96     \\
Max near mem BW (GB/s)             & 138     & 46     & 56     \\
Max far mem BW (GB/s)              & 7.45/2.25 & 6.8/1.85 & 36/36 \\
Near memory latency (ns)   & 80      & 80     & 95     \\
Far memory latency (ns)    & 150 - 250 & 150 - 250 & 145 \\
\bottomrule
\end{tabular}
\vspace{1ex}
\caption{Specifications of machines used in our evaluation.}
\label{tab:machine_spec}
\vspace{-5ex}
\end{table}

\subsection{Experimental Setup}

\noindentparagraph{Hardware setup.}
Table~\ref{tab:machine_spec} shows the hardware specifications of the three machines used in our evaluation. \texttt{pmem-large} and \texttt{pmem-small} have Intel Optane DC Persistent Memory DIMMs which form the slower tier, whereas the \texttt{NUMA} machine (provided by Cloudlab~\cite{cloudlab}) is used for emulating CXL memory. Notably, \texttt{pmem-small} has $3\times$ less DRAM bandwidth compared to \texttt{pmem-large}. Unless mentioned otherwise, we run our experiments on \texttt{pmem-large}.

\noindentparagraph{Workloads.}
For our evaluation, we select a set of 8 representative and diverse workloads.
Specifically, we employ three graph processing algorithms from GapBS~\cite{beamer2015gap}, Graph500 \cite{murphy2010introducing}, an HPC workload (XSBench \cite{tramm2014xsbench}), an in-memory database (Silo \cite{tu2013speedy}), an in-memory index lookup (Btree \cite{reto2020btree}), and a popular memory intensive micro-benchmark, GUPS \cite{plimpton2006simple}. 
Table~\ref{tab:benchmark_specs} contains additional information about these workloads, including the resident set size (RSS) and the different inputs used (if any) for each workload.
This set of workloads have been widely used to evaluate tiered memory systems in many prior works \cite{hemem2021, memtis2023, johnnyCache2023,autotiering2017}.
When measuring execution time, we make sure to only consider relevant applications phases (e.g., graph construction/traversal, query execution), ignoring any potential reading/initialization phases that are not part of the actual kernel (e.g., disk I/O).
We run these workloads with a thread count large enough to just saturate the memory bandwidth of each system, i.e., 12 threads for \texttt{pmem-large} and \texttt{NUMA}, and 4 threads for \texttt{pmem-small}.

\begin{table}[t]
\scriptsize
\centering
\begin{tabular}{@{}llll@{}}
\toprule
\textbf{Workload} & \textbf{Inputs} & \textbf{RSS} & \textbf{Description} \\ \midrule
\multirow{2}{*}{GapBS-BC \cite{beamer2015gap}} & kronecker  & 78.13 & Compute the measure of centrality  \\
 & twitter & 13.08 &  in a graph based on shortest paths. \\ \midrule
\multirow{2}{*}{GapBS-PR \cite{beamer2015gap}} & kronecker  & 71.29 & \multirow{2}{*}{Compute the PageRank score of a graph} \\
 & twitter & 12.32 & \\ \midrule
\multirow{2}{*}{GapBS-CC \cite{beamer2015gap}} & kronecker  & 69.29 & Compute connected components of \\
 & twitter & 12.09 &  a graph using (Shiloach-Vishkin) \\ \midrule
\multirow{2}{*}{Silo \cite{tu2013speedy}} & TPC-C & 75.68 & \multirow{2}{*}{In-memory transactional database} \\
 & YCSB-C & 71.40 & \\ \midrule
Btree \cite{reto2020btree} & - & 12.13 &  In-memory index lookup benchmark \\ \midrule
XSBench \cite{tramm2014xsbench} & - & 64.97 & Compute kernel of the MCNP algorithm \\ \midrule
GUPS \cite{plimpton2006simple} & 8 GiB hot & 64.03 & Random accesses with dynamic hotset \\ \midrule
Graph500 \cite{murphy2010introducing} & kronecker & 34.13 & Construction and BFS of large graphs \\
\bottomrule
\end{tabular}
\vspace{1ex}
\caption{Workload Characteristics. RSS is in GiB.}
\label{tab:benchmark_specs}
\vspace{-5ex}
\end{table}

\noindentparagraph{Tiering Configuration.} Similar to Memtis~\cite{memtis2023}, we configure the ratio of fast to slow tier memory size by setting the fast tier size to the corresponding proportion of the workload resident set size (RSS). Unless otherwise noted, experiments maintain a 1:8 memory size ratio. For instance, for GUPS whose RSS is ~64 GB, we set the fast tier size to 7.11 GB (11\%).
In HeMem, we set the fast tier size as a macro in a header file. In HMSDK, we run a background \textit{memchew} workload that reserves the desired portion of the fast tier, but does not perform any memory operations.

\noindentparagraph{Optimizer Configuration.}
We configure SMAC to run with a budget of $100$ iterations, using the first $20$ as an initial exploration phase; we also force the optimizer to pick a random configuration with $20\%$ probability.
Similar setups have been used in prior works, for more complex systems~\cite{kanellis2022llamatune}.
HeMem exposes its knobs as macros in the library. At every iteration, in order to deploy the newly-suggested system configuration, the optimizer modifies the values of these macros and re-compiles the library before starting the execution.
HMSDK can be configured by passing a JSON file as input.

\subsection{Performance benefit of parameter tuning}
\label{sec:eval-scailp-hemem}

\begin{figure}[t]
    \setlength{\abovecaptionskip}{1ex}
    \centering
    \includegraphics[width=\linewidth]{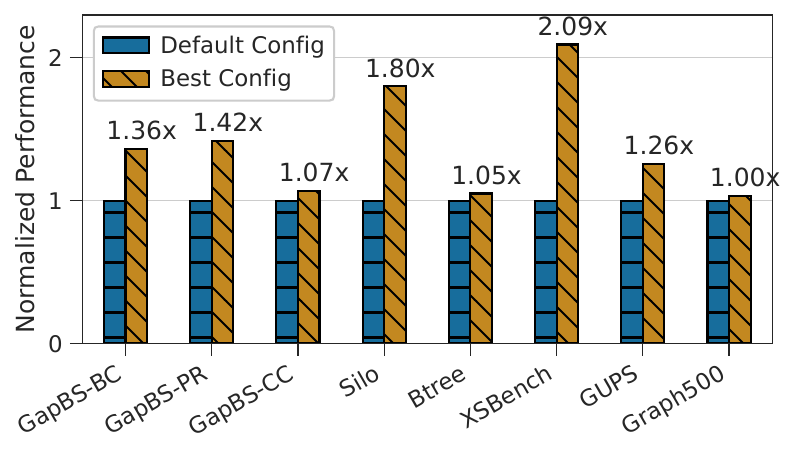}
    \caption{Performance improvement of best HeMem configuration found over default for all workloads.}
    \label{fig:hemem-improv-scailp}
    \vspace{-2ex}
\end{figure}

We begin by comparing the performance of the best knob configuration found by the optimizer against the performance with the default HeMem configuration. We consider each workload in our benchmark set and also analyze why the optimizer configuration is better by relating to the workload behavior. 
Figure~\ref{fig:hemem-improv-scailp} shows the results. For almost all workloads barring Graph500, the optimizer identifies parameter values that provide superior performance, by $1.07-2.09x$.

Below, we briefly discuss the differences in parameter values and the reasons for these improvements, while Table~\ref{tab:summary_performance} summarizes our findings.

\begin{figure}[t]
    \centering
    \setlength{\abovecaptionskip}{1ex}
    \begin{minipage}{0.46\linewidth}
        \includegraphics[width=\linewidth]{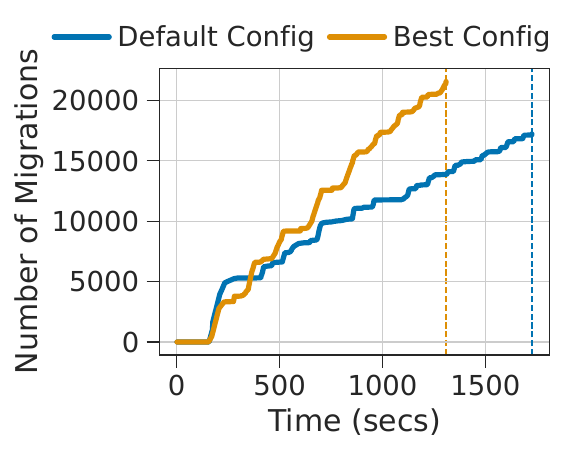}
        \caption{\textit{GapBS-BC}: Number of migrations over time.}
        \label{fig:bc-migrations}
    \end{minipage}
    \hspace{1em}
    \begin{minipage}{0.47\linewidth}
        \includegraphics[width=\linewidth]{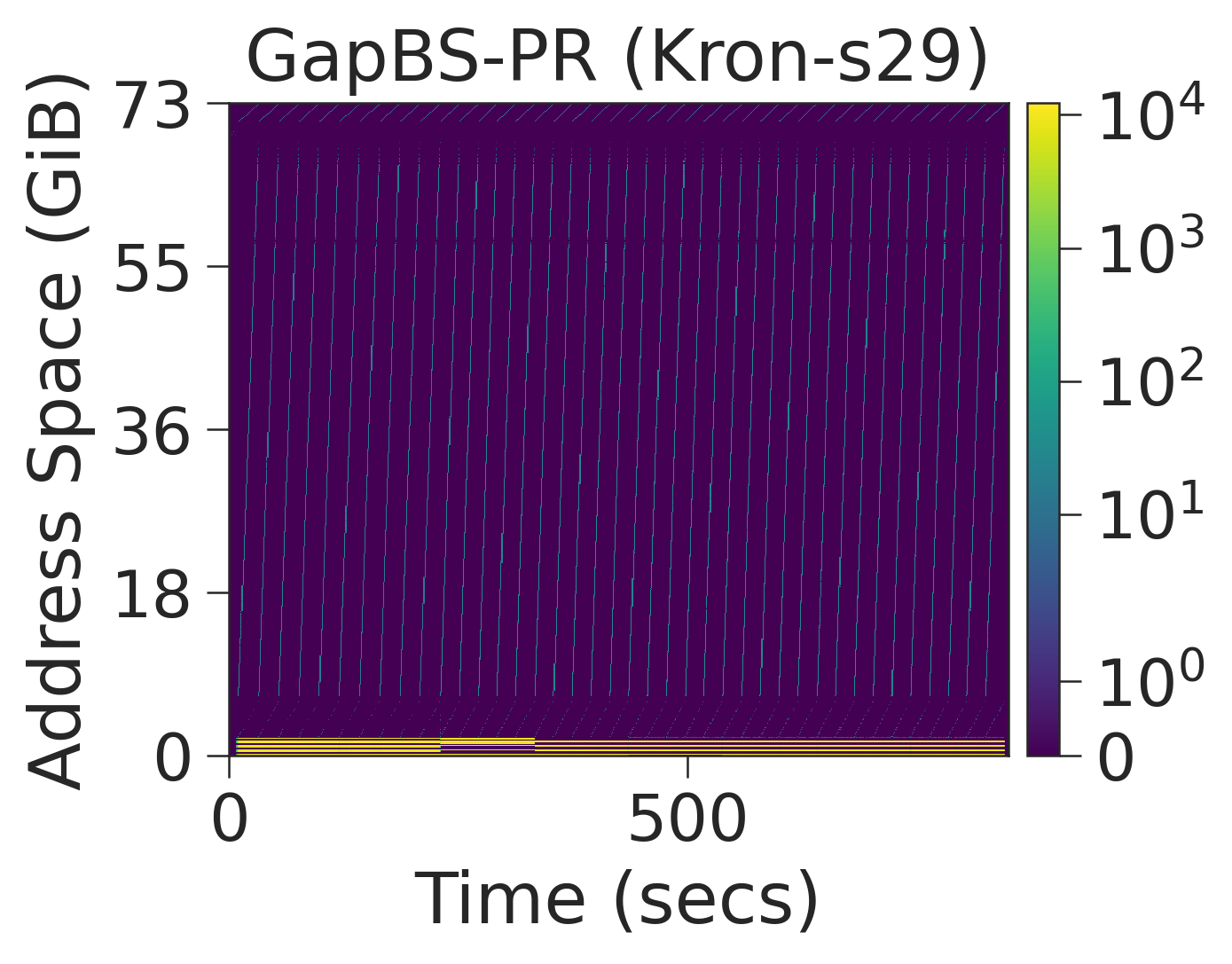}
        \caption{\textit{GapBS-PR}: Memory access pattern over time.}
        \label{fig:pr-memory-access}
    \end{minipage}
    \vspace{-2ex}
\end{figure}

\smalltitle{GapBS-BC (kronecker graph): } We run Betweenness Centrality (BC) from the GapBS suite with a synthetic kronecker graph with 1 million vertices. We find that the best configuration found by the optimizer outperforms the default by about 1.36x. The optimizer tunes the hotness thresholds and cooling threshold which results in a more accurate classification of hot and cold pages, and timely migrations. Figure~\ref{fig:bc-migrations} shows how the number of migrations increases over time for the default and best configurations. With the default configuration (\texttt{read\_hot\_threshold} of 8), HeMem takes longer to identify all the hot pages, even fails to identify some of them. With the best configuration (\texttt{read\_hot\_threshold} of 4 or 5), HeMem identifies most of the hot pages quickly at the beginning of every iteration and promotes them to fast tier (each step in the graph corresponds to the beginning of an iteration). We also find that tuning the \texttt{cooling\_threshold} is important to avoid frequent cooling which could trigger demotions of actually hot pages. 

\smalltitle{GapBS-PR, CC (kronecker graph): } With PageRank (PR) and Connected Components (CC), we find that a small set of pages are hot while the rest of the pages are cold and accessed infrequently. As seen in Figure~\ref{fig:pr-memory-access}, PR exhibits a streaming memory access pattern for the cold pages. The optimizer learns that PR does not benefit from migrations of these cold pages as there is little to no reuse. On the other hand, the default configuration keeps migrating pages throughout the experiment which eats into the memory bandwidth and stalls many write accesses, slowing down the application. Similar to PR above, CC also has a streaming access pattern and doesn't benefit from migrations. The best-performing configuration is similar, in that it avoids any wasteful migrations. 

\vspace{-1ex}
\begin{mdframed}[linecolor=black,linewidth=1pt,innerleftmargin=8pt,innerrightmargin=8pt,innertopmargin=4pt,innerbottommargin=4pt, skipabove=8pt]
\textbf{Takeaway: }\textit{Knowing the workload behavior and possibly the graph structure, in which vertices are popular while others are not, can help an ideal tiering system place appropriate pages in the faster tier and avoid all migrations.}
\end{mdframed}
\vspace{1ex}

\begin{table}[t]
\centering
\fontsize{6.6}{8.2}\selectfont
\begin{tabular}{@{}lll@{}}
\toprule
        & \textbf{Reason(s) for improvement}  & \textbf{Important knobs}       \\ \midrule
BC      & \begin{tabular}[c]{@{}l@{}}Accurate classification of\\ hot/cold pages, timely migrations\end{tabular}                                      & \begin{tabular}[c]{@{}l@{}}read\_hot\_threshold,\\ sampling\_period, cooling\_threshold\end{tabular} \\ \cmidrule{1-3}
PR,CC   & Eliminate unnecessary migrations                                                                                                            & \begin{tabular}[c]{@{}l@{}}read\_hot\_threshold, write\_hot\_threshold,\\ cooling\_threshold\end{tabular}                   \\\cmidrule{1-3}
Silo    & \begin{tabular}[c]{@{}l@{}}Better hot pages identification,\\ reduce migrations of warm pages,\\ lower write sampling overhead\end{tabular} & \begin{tabular}[c]{@{}l@{}}read\_hot\_threshold, sampling\_period,\\ cooling\_pages\end{tabular}                            \\\cmidrule{1-3}
Btree   & \begin{tabular}[c]{@{}l@{}}Decrease importance\\  of write-heavy pages\end{tabular}                                                         & \begin{tabular}[c]{@{}l@{}}write\_hot\_threshold,\\ write\_sampling\_period\end{tabular}                                    \\\cmidrule{1-3}
XSBench & Eliminate warm page migrations                                                                                                              & \begin{tabular}[c]{@{}l@{}}read\_hot\_threshold, write\_hot\_threshold,\\ cooling\_threshold\end{tabular}                   \\\cmidrule{1-3}
GUPS    & \begin{tabular}[c]{@{}l@{}}Increase sampling accuracy,\\ faster detection of hot pages\end{tabular}                                         & \begin{tabular}[c]{@{}l@{}}read\_hot\_threshold, sampling\_period\\ write\_hot\_threshold, sampling\_period\end{tabular}    \\ \bottomrule
\end{tabular}
\vspace{1ex}
\caption{Summary of performance improvement from parameter tuning on pmem-large for different workloads}
\label{tab:summary_performance}
\vspace{-5ex}
\end{table}

\smalltitle{Silo (YCSB-C):} Silo is an in-memory transactional database.
Similar to Memtis~\cite{memtis2023}, we run Silo with YCSB-C read-only workload. With this workload, about 1\% (700MB) of the pages of Silo are extremely hot, while about 20\% of the pages are warm.
The optimizer generates configurations that perform better than the default. Our analysis reveales three main reasons: (1) the best configuration identifies hot pages more accurately and quickly by reducing \texttt{sampling\_period} and reducing \texttt{read\_hot\_threshold}), (2) it reduces migrations of moderately warm pages by cooling all pages at the same time rather than in batches (increases \texttt{cooling\_pages}), and (3) reduces write sampling overheads, which makes sense since YCSB-C is a read-only workload.

The important knob analysis concluded that value of \texttt{cooling\_pages} plays an important role in Silo's performance. This knob is one of the aforementioned hidden knobs that are usually neglected, highlighting the need to consider all possible knobs.

\smalltitle{Btree:} We populate an in-memory Btree with $600$M key-value pairs and perform $750$M lookups. This workload has $2$ phases: an initialization phase where keys are inserted into the Btree, and a lookup phase, where uniform random keys are queried.
The initialization phase is write-heavy as new elements are created and added to the tree. Note that the tree might be adjusted after every insert to meet Btree properties. With the default configuration, HeMem starts migrating pages that are being written to. Since HeMem uses a low \texttt{write\_hot\_threshold}, a lot of pages get hot and are migrated. In fact, about $16,000$ migrations out of the $18,000$ migrations happen during this initialization phase. The optimizer realizes that the hot pages in the initialization phase will not remain hot during the lookup phase. Thus, it generates configurations with higher \texttt{write\_hot\_threshold}, so very few write-heavy pages are migrated during initialization. This creates room for read-hot pages such as those holding the high level nodes in the fast tier. This in turn reduces the latency of lookup queries.

\vspace{-1ex}
\begin{mdframed}[linecolor=black,linewidth=1pt,innerleftmargin=8pt,innerrightmargin=8pt,innertopmargin=4pt,innerbottommargin=4pt, skipabove=8pt]
\textbf{Takeaway: }\textit{Identifying that program phases are read-only (write-heavy) can help an ideal tiering system adjust the appropriate read hot (write hot) thresholds and minimize migrations.}
\end{mdframed}
\vspace{1.5ex}

\begin{figure}[t]
    \centering
    \setlength{\abovecaptionskip}{1ex}
    \begin{minipage}{0.47\linewidth}
        \includegraphics[width=\linewidth]{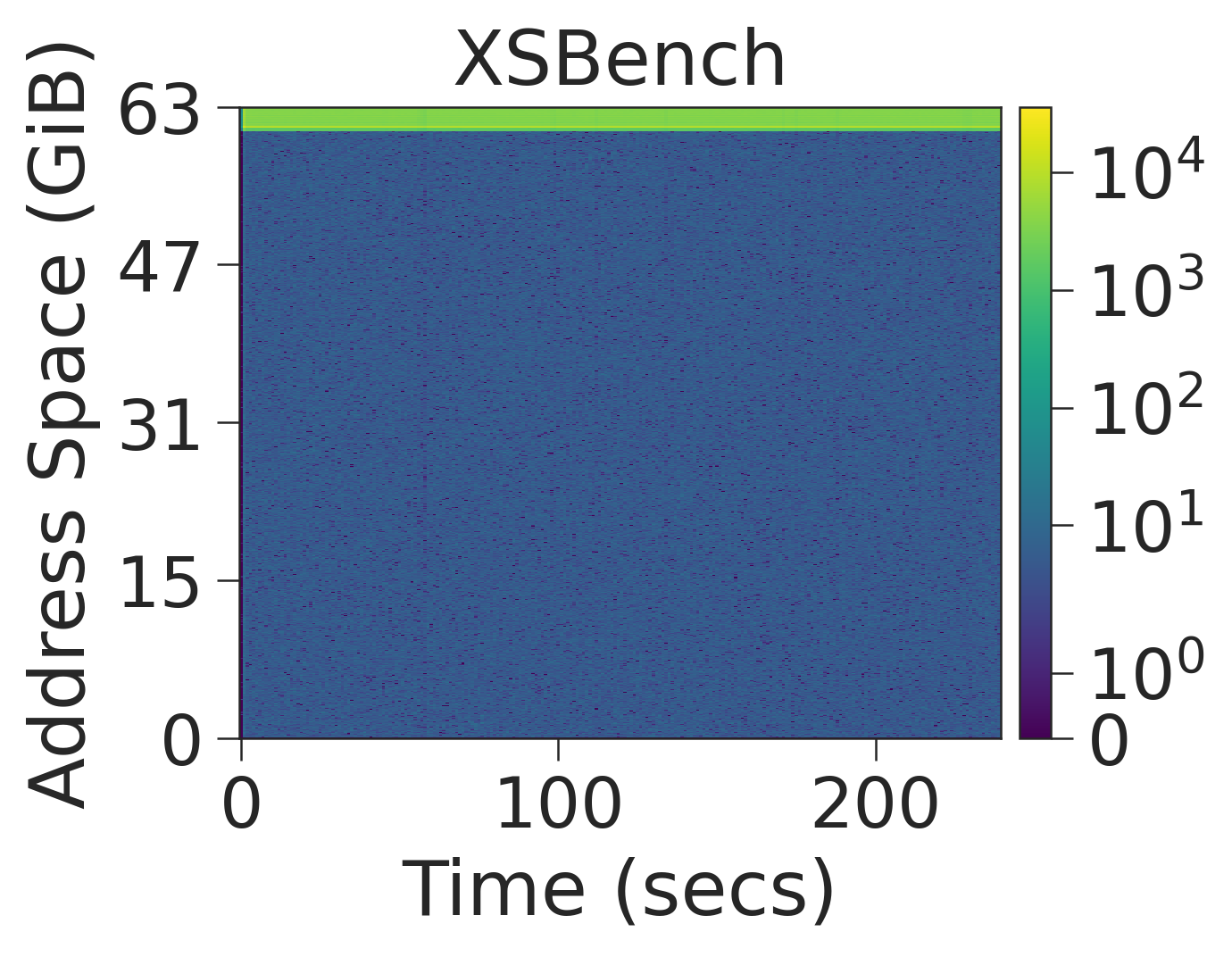}
        \vspace{-2ex}
        \caption{\textit{XSBench}: Memory access pattern over time.}
        \label{fig:xsbench-memory-access}
    \end{minipage}
    \hspace{.5em}
    \begin{minipage}{0.47\linewidth}
        \includegraphics[width=\linewidth]{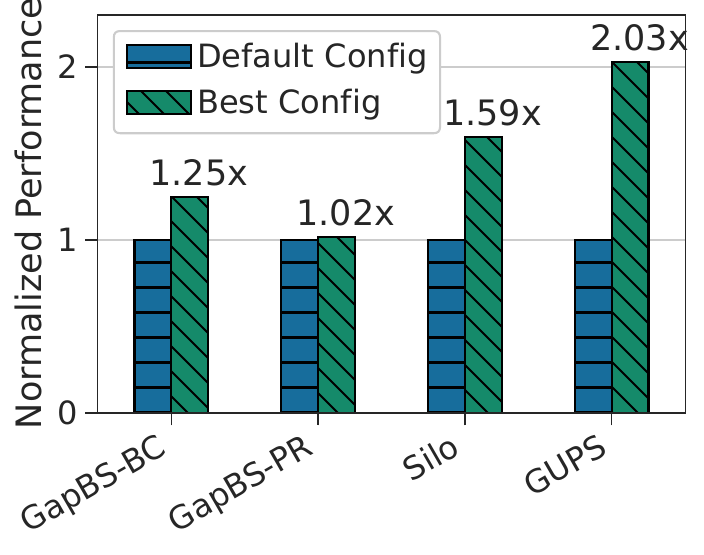}
        \vspace{-2ex}
        \caption{Performance gains (vs default) for \texttt{pmem-small}.}
        \label{fig:hemem-improv-josepm}
    \end{minipage}
    \vspace{-2ex}
\end{figure}

\smalltitle{XSBench:} Figure~\ref{fig:xsbench-memory-access} shows the access heatmap of XSBench Monte Carlo algorithm. XSBench has a small set of pages that are frequently accessed (greenish-yellow line at the top). Rest of the pages (blue area) are randomly accessed and have very similar access counts. Thus, it is important to keep the hot pages in the fast tier; it does not matter which of the remaining pages are also in the fast tier. The optimizer learns this and converges to configs that keep the hot pages in fast tier, avoiding wasteful migrations of the randomly accessed pages. To do so, it sets the \texttt{read\_hot\_threshold} and \texttt{write\_hot\_threshold} higher than the \texttt{cooling\_threshold}, to not trigger any migrations.

\smalltitle{GUPS:} We evaluate GUPS micro-benchmark with a skewed access distribution similar to prior work~\cite{hemem2021,johnnyCache2023}. We configure GUPS with a hotset of 8GiB and a total memory footprint of 64GiB. The hotset moves after half of the updates are performed. For our 1:8 memory configuration, the hotset spills out of the fast tier. With the default configuration, HeMem keeps shuffling hot pages between the tiers because of low sampling accuracy. The optimizer recognizes this issue and increases the sampling frequency resulting in more accurate access sampling. This does not only improve hot page identification, but it also reduces the wasteful migrations opening up more bandwidth for the application.

\vspace{-1ex}
\begin{mdframed}[linecolor=black,linewidth=1pt,innerleftmargin=8pt,innerrightmargin=8pt,innertopmargin=4pt,innerbottommargin=4pt, skipabove=8pt]
\textbf{Takeaway: }\textit{Although increasing sampling frequency incurs higher overheads, some workloads can benefit more from the increased accuracy. So, constraining sampling overheads might not be a good idea for some applications.}
\end{mdframed}
\vspace{1ex}

\paragraph{Validating results on pmem-small:}
We run similar experiments to the above on our smaller pmem-small machine to verify that (1) performance gains are possible when switching to different hardware, and (2) the optimizer can identify similar best-performing configurations. Since pmem-small has smaller DRAM bandwidth, we run the same applications with fewer threads and smaller inputs. Figure~\ref{fig:hemem-improv-josepm} shows the performance difference between the default configuration of HeMem and the best configurations identified by the optimizer. Overall, the results are very similar to those obtained on pmem-large, corroborating the results discussed above.

\begin{figure}[t]
    \centering
    \setlength{\abovecaptionskip}{1ex}
    \includegraphics[width=\linewidth]{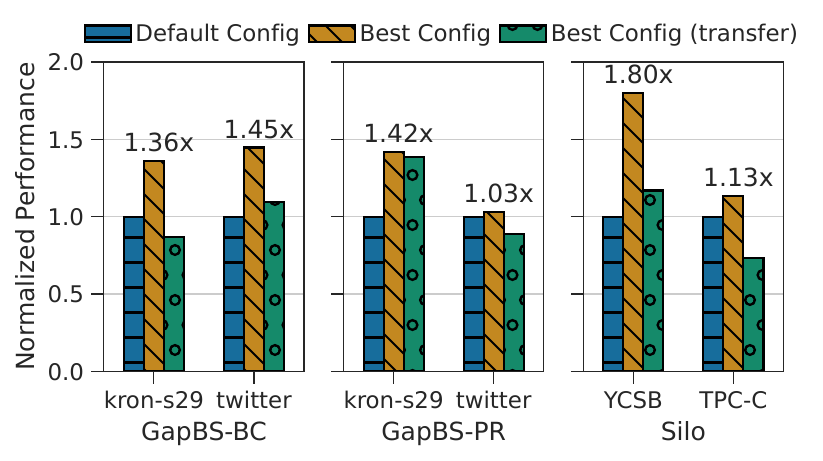}
    \caption{Performance of HeMem's best configuration compared to default, for two application inputs. We also show how each best configuration performed when HeMem used it while running on the alternative input (i.e., \textit{transfer}).}
    \label{fig:hemem-improv-scailp-inputs}
    \vspace{-2ex}
\end{figure}

\subsection{Tuning with different application inputs}
Application memory access patterns depend heavily on the input datasets/workload. For instance, the memory access pattern for graph processing applications could be very different for different graphs. Similarly, key-values (KV) stores exhibit different memory characteristics when running against different workloads (e.g. TPC-C versus YCSB-C). Since application behavior changes across inputs, the best-performing tiering knob values might also be different for different inputs. To answer this question, we ran our optimizer pipeline for graph and DB workloads with different inputs.

To understand if knob values are dependent on inputs, we use the best knob values obtained for one input with the other input. For example, we ran GapBS-BC on a twitter graph using the best configuration obtained for the kronecker graph. Figure~\ref{fig:hemem-improv-scailp-inputs} shows the results. As expected, in most cases, the best configuration generated for one input does not perform well for the other input. In fact, we observe that the performance is worse than default in most cases. 

Below, we explain some of the key differences between the best configurations we found for the different inputs.

\begin{figure}[t]
    \centering
    \setlength{\abovecaptionskip}{1ex}
    \begin{minipage}{0.49\linewidth}
        \includegraphics[width=\linewidth]{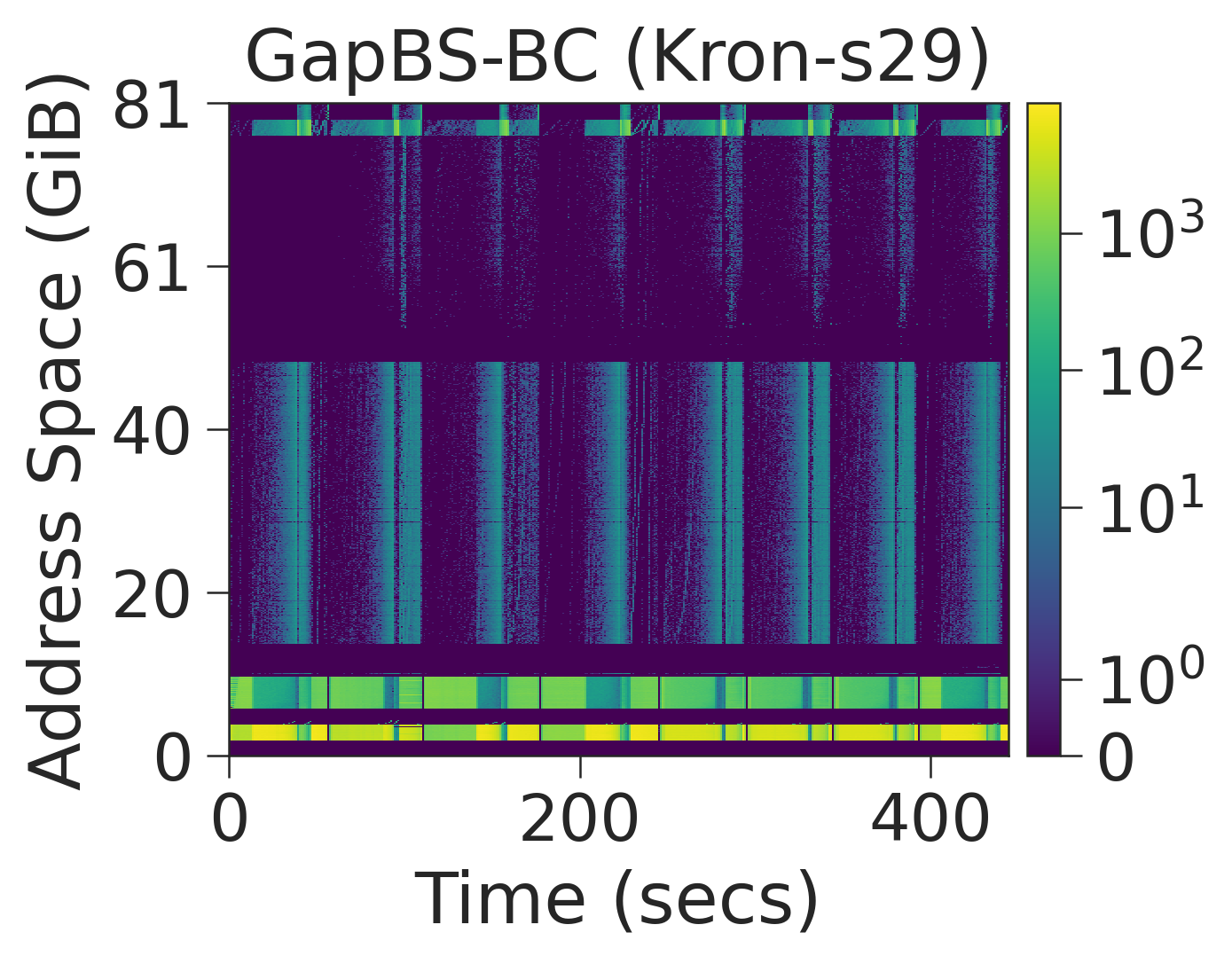}
    \end{minipage}
    \begin{minipage}{0.49\linewidth}
        \includegraphics[width=\linewidth]{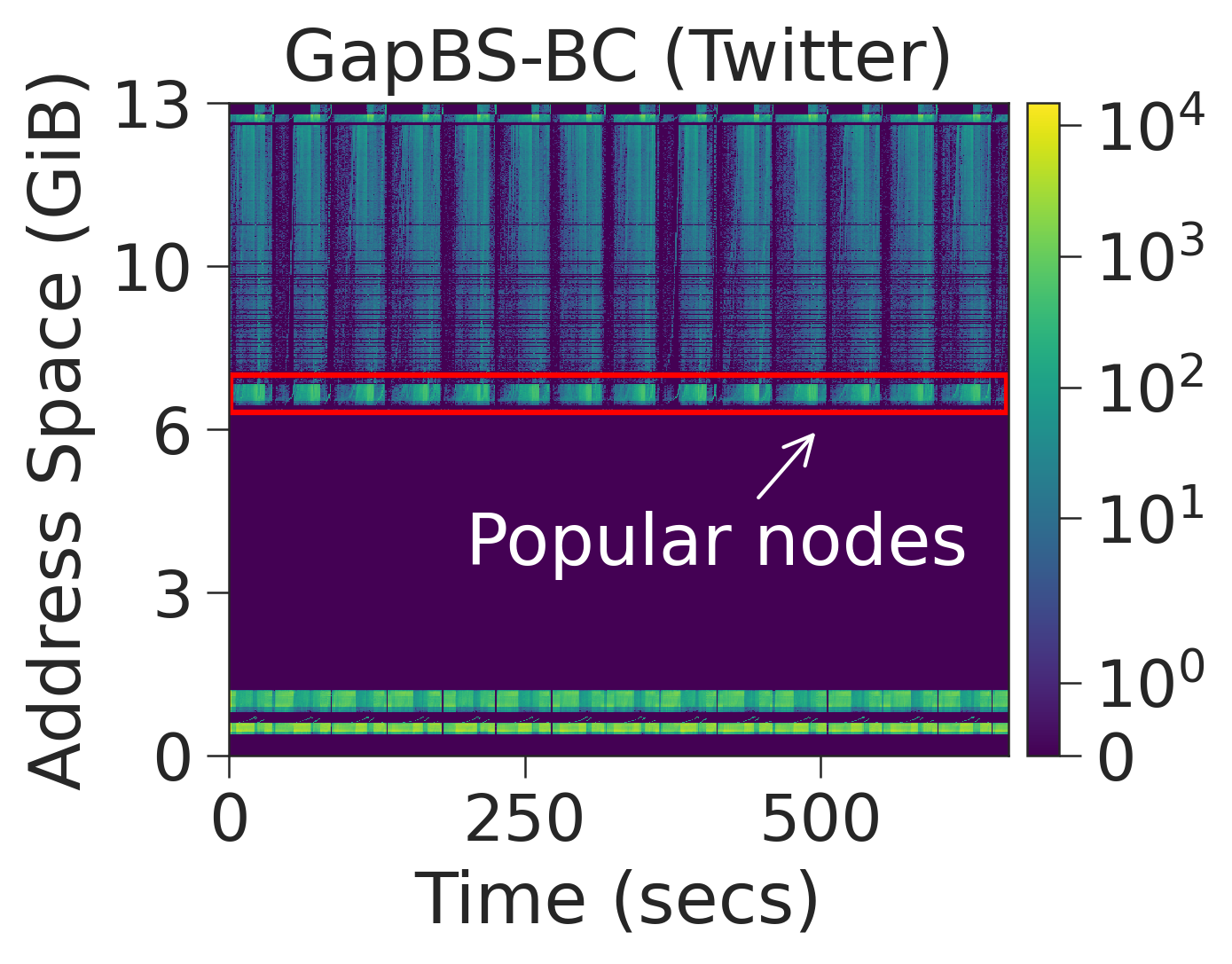}
    \end{minipage}
    \caption{\textit{GapBS-BC} memory access pattern over time, for Kronecker-s29 (left) and Twitter (right) input graphs.}
    \label{fig:bc-access-pattern}
    \vspace{-2ex}
\end{figure}

\smalltitle{Gap-BC, PR:} We compare the results of optimization runs for BC and PR with different inputs: kronecker and Twitter graphs. The memory access patterns is similar for both the graphs except for one difference (see Figure~\ref{fig:bc-access-pattern}). With the twitter graph, there are a handful nodes corresponding to influential profiles that are very popular. These popular nodes are present on a small set of pages as shown in Figure~\ref{fig:bc-access-pattern}. This results in frequent accesses to these pages whereas in kronecker, all pages receive almost equal number of accesses. The optimizer successfully identifies this difference. For the Twitter graph, the optimizer selects configuration which allow the the hot pages corresponding to the popular nodes to be migrated early. It does so by lowering \texttt{write\_sampling\_period} and the \texttt{write\_hot\_threshold} allowing for accurate and faster promotion of the popular pages.

\smalltitle{Silo:} We compare the best configurations for Silo with YCSB-C and TPC-C workloads. YCSB-C and TPC-C are very different workloads. While YCSB-C is a read-only workload with Zipfian access distribution, TPC-C is an insert-heavy workload that mostly reads new data, meaning that pages are hot when they are updated but become cold quickly as new entries are inserted into the database.
As discussed earlier, for YCSB-C, the optimizer prioritizes read sampling parameters and de-prioritizes write sampling.
On the other hand, for TPC-C, the optimizer recognizes the importance of writes. It arrives at configurations that promotes newly written pages to the fast tier for a short time and demotes them to slow tier as soon as they lose their hotness. The best-performing configurations (1) prioritize written pages by lowering write hot threshold and decreasing write sampling period, (2) invoke cooling frequently because hot set changes quickly, and (3) run the migration thread frequently by setting the migration period to the lowest value.

\vspace{-1ex}
\begin{mdframed}[linecolor=black,linewidth=1pt,innerleftmargin=8pt,innerrightmargin=8pt,innertopmargin=4pt,innerbottommargin=4pt, skipabove=8pt]
\textbf{Takeaway: }\textit{Application behavior is tied to the inputs they are processing. Thus, it is not sufficient to find just a single set of knob values for an application. System administrators need to consider the input properties along with the application when configuring knob values.}
\end{mdframed}
\vspace{1ex}


\begin{figure}[t]
    \setlength{\abovecaptionskip}{1ex}
    \centering
    \begin{minipage}{0.47\linewidth}
    \centering
    \includegraphics[width=\linewidth]{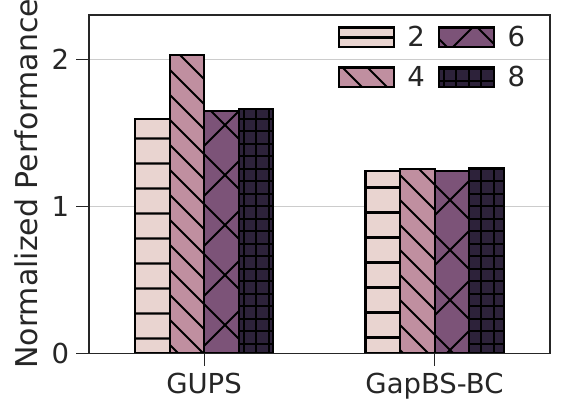}
    (a) Varying Thread Count
    \end{minipage}
    \begin{minipage}{0.47\linewidth}
    \centering
    \includegraphics[width=\linewidth]{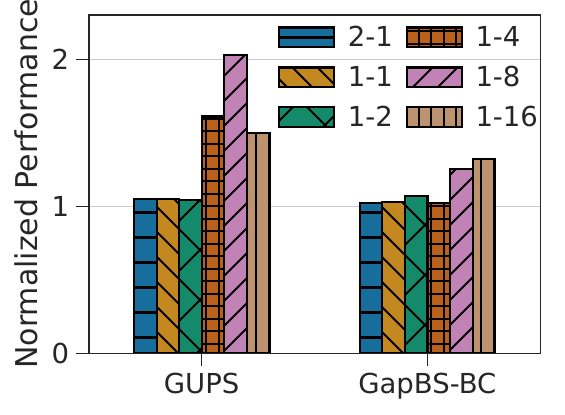}
    (b) Varying Memory Ratios
    \end{minipage}
    \caption{Performance gains (over default) of HeMem's best configuration found, for varying (a) number of threads and (b) memory size ratios.}
    \label{fig:varying-system-configuration}
    \vspace{-2ex}
\end{figure}

\subsection{Tuning for different system configurations}

Memory tiering parameter values could depend on the properties of the underlying system. This includes various factors, significant among which are the number of cores, memory sizes and memory bandwidth. In this section, we look at the relationship between knob values and the aforementioned hardware characteristics.

\subsubsection{Different number of threads.}
\label{sec:varying-num-threads}

Programmers often configure the number of threads for an application based on the number of cores and the memory bandwidth available. Increasing the number of threads increases memory traffic which results in more accesses per page in shorter time period. Therefore, tiering engines need to have different thresholds and periods based on the number of threads.

Using two workloads, GUPS and BC (twitter graph), we perform optimization runs varying the number of threads. Figure~\ref{fig:varying-system-configuration}a shows the performance improvement we achieve by changing knob values for different number of threads, on \texttt{pmem-small}. We see consistent performance improvement for all thread counts for both workloads. We see that the best knob values are \textit{very different} for different thread counts. No single configuration performs well for all thread counts.

Through our analysis, we find that the \texttt{read\_hot\_threshold} is the most important knob and its value increases with number of threads, as expected. However, we find that it is not sufficient to just tune this single parameter when changing the number of threads. Other knobs such as the \texttt{migration\_period} and \texttt{cooling\_threshold} also have to be tuned accordingly.

\subsubsection{Different memory size ratios.}
To understand the relationship between knob values and memory tiering size configurations, we run optimization on the knobs for different fast-slow tier memory size ratios. Recall that the memory sizes of the fast and slow tier are configured based on the measured RSS of the workload.

Figure~\ref{fig:varying-system-configuration}b shows the performance improvement over default knob setting for different memory ratios on \texttt{pmem-small}. The key observation here is that the HeMem configuration becomes more important for smaller fast tier sizes, where the entire hot set might not be able to fit in fast tier (e.g., 1:16, 1:8). Indeed, we find that the default configuration performs poorly in such settings. The optimizer is able to identify knob values that result in more accurate page classification and timely migrations. For instance, the optimizer sets the hot thresholds higher in the 1:16 and 1:8 tiering configurations and sets them lower in the 1:2, 1:1 and 2:1 configurations. As a result, only the extremely hot pages are migrated in the 1:16 and 1:8 systems whereas any page sampled/accessed is promoted in the larger fast tier configurations.

\begin{figure}[t]
    \centering
    \setlength{\abovecaptionskip}{0pt}
    \includegraphics[width=0.97\linewidth]{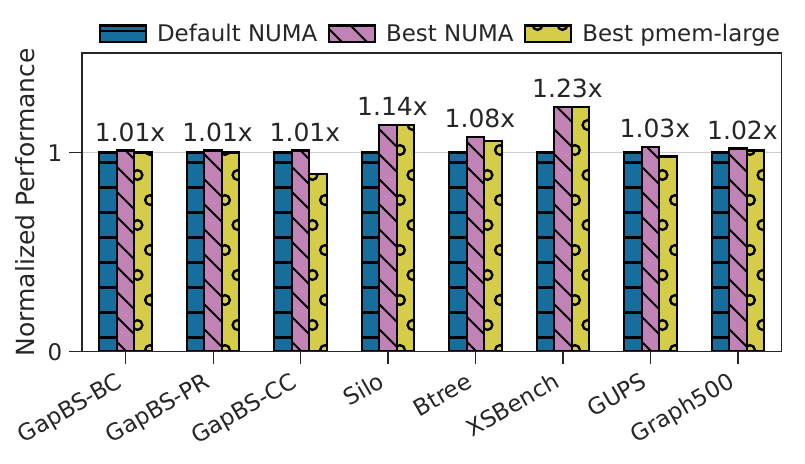}
    \caption{Performance gains (over default) of HeMem's \textit{best} configuration for NUMA machine (annotated). We also show how \texttt{pmem-large} best config performed when deployed on NUMA machine.}
    \label{fig:scailp-vs-numa}
    \vspace{-2ex}
\end{figure}

\subsubsection{Different memory bandwidth.}
Studies on CXL show that CXL memories can offer latency and bandwidth comparable to NUMA~\cite{cxlperf2023}. Therefore, we emulate CXL memory using a NUMA machine wherein the local node memory represents fast tier and remote node memory emulates the slow tier. Table~\ref{tab:machine_spec} shows the latency and bandwidth of local and remote NUMA memories. 

Figure~\ref{fig:scailp-vs-numa} shows the performance improvement achievable by tuning knobs of HeMem while running on the NUMA machine.
We observe that the performance gains are mostly modest.
The explanation is straight-forward: since latencies and bandwidth are similar between the tiers, there is little room for performance improvement. Second, migrations have little to no interference with application accesses because of high available bandwidth. Therefore, the penalty for potential wasteful migrations is negligible, so tiering systems do not need to be systematically tuned to perform well.

Interestingly, we also observe that best performing configurations on pmem-large mostly perform well even on the NUMA machine (e.g. Silo, Btree, XSBench). It might therefore be possible to transfer the best-performing configuration from one machine to another machine for some workloads.

\begin{figure}[t]
    \centering
    \setlength{\abovecaptionskip}{0pt}
    \includegraphics[width=\linewidth]{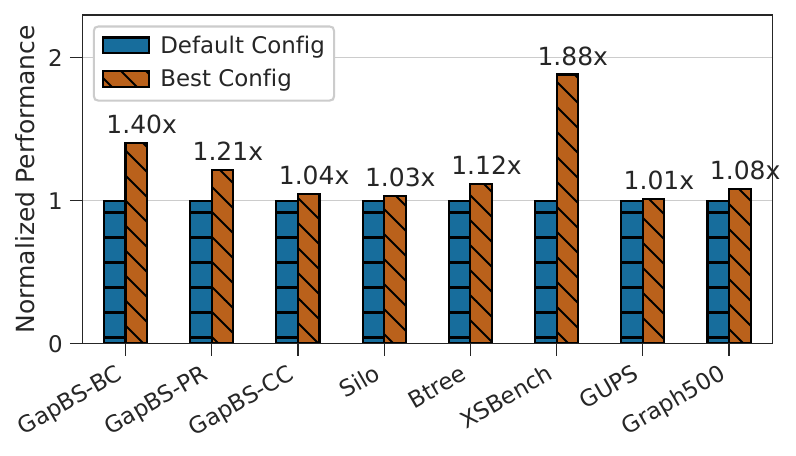}
    \caption{Performance of best HMSDK configuration found (over default) for all workloads on NUMA machine.}
    \label{fig:hmsdk}
    \vspace{-2ex}
\end{figure}

\subsection{Tuning other memory tiering systems}
\label{sec:hmsdk_tuning}
As discussed in Section~\ref{sec:background}, almost all of the existing memory tiering systems use heuristics and have knobs of different kinds. In the previous sections, we saw how changing tiering parameter values improves HeMem's performance under various scenarios.
To explore whether similar can be extended to other tiering systems, we also experiment with Heterogeneous Memory Software Development Kit (HMSDK)~\cite{hmsdk2024}. 
HMSDK is an open-source memory tiering system developed by SK-Hynix that uses DAMON~\cite{damon2019} to identify hot and cold pages in memory. DAMON is a data access monitoring framework subsystem for the Linux kernel that scans page tables periodically. To reduce the overheads, DAMON divides the address space into a number of regions and samples page table entries from each region. HMSDK, like HeMem, has several knobs related to access monitoring, hot/cold page classification and migrations.

We setup HMSDK on the NUMA machine emulating CXL and run all workloads with $12$ threads. Figure~\ref{fig:hmsdk} shows the performance improvement achieved by changing various knob values in HMSDK. Again, we observe significant gains for some workloads and modest gains with others. Below, we discuss some key differences between the default knob values and the best knob values that result in better performance.

\smalltitle{GapBS-PR and Btree:} With the default knob values, HMSDK is unable to correctly identify hot pages and promote them. The optimizer improves page access monitoring by increasing both the number of DAMON regions (\texttt{nr\_regions}) and the scanning frequency (\texttt{intervals:sample\_us}). This leads to migrations inside the correct hot regions and reduces erroneous or unnecessary migrations, thereby improving performance.

\smalltitle{XSBench:} As discussed previously, XSBench has a small set of hot pages which should be placed in near memory while the rest of the pages have very similar access frequency. Similar to HeMem, we observe that HMSDK by default performs a large number of unnecessary migrations (about 10 million pages). The optimizer identifies knob settings that eliminate migrations of such pages with similar access frequency, which improves performance significantly.

\begin{figure}[t]
    \centering
    \setlength{\abovecaptionskip}{1ex}
    \begin{minipage}{0.49\linewidth}
        \includegraphics[width=\linewidth]{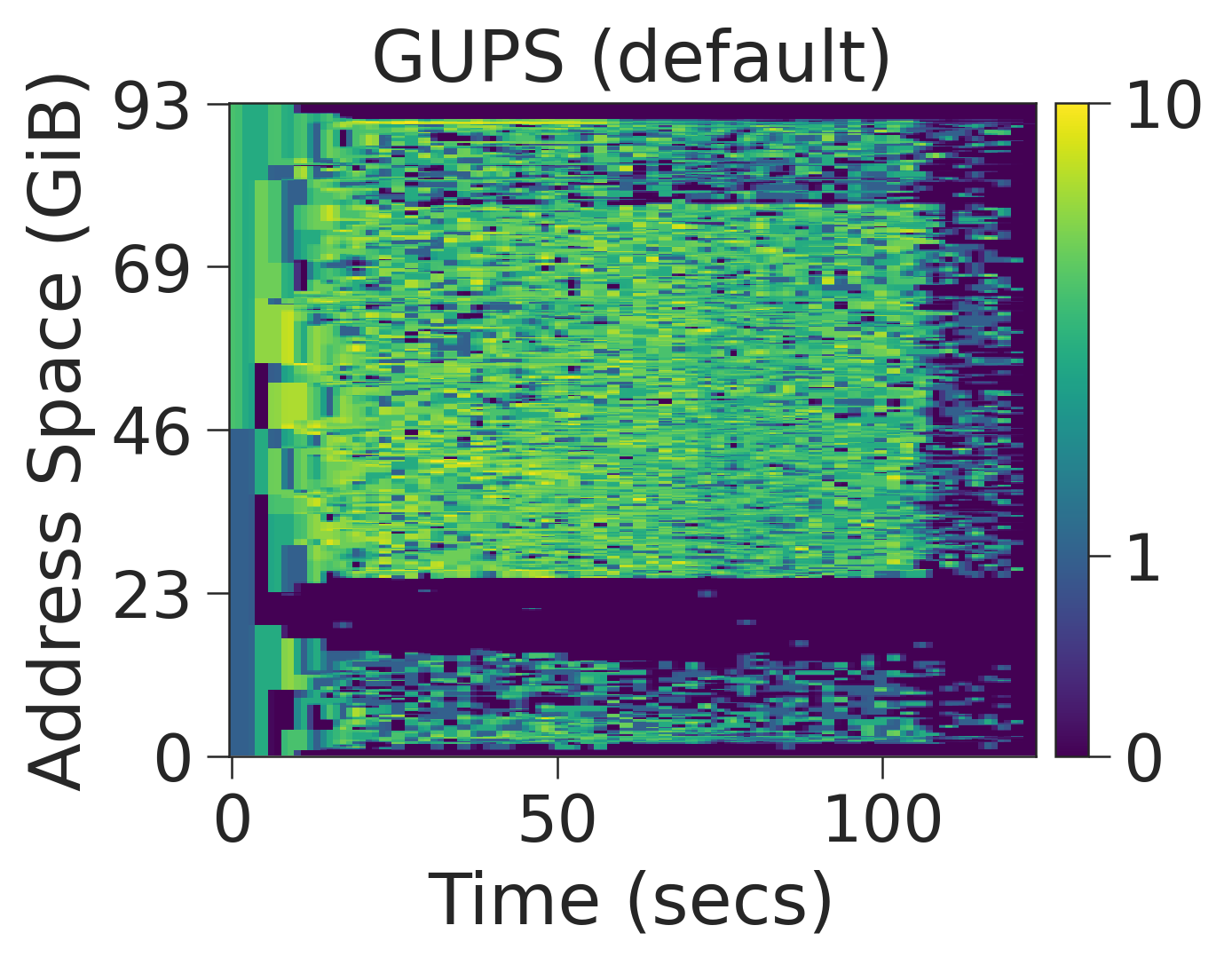}
    \end{minipage}
    \begin{minipage}{0.49\linewidth}
        \includegraphics[width=\linewidth]{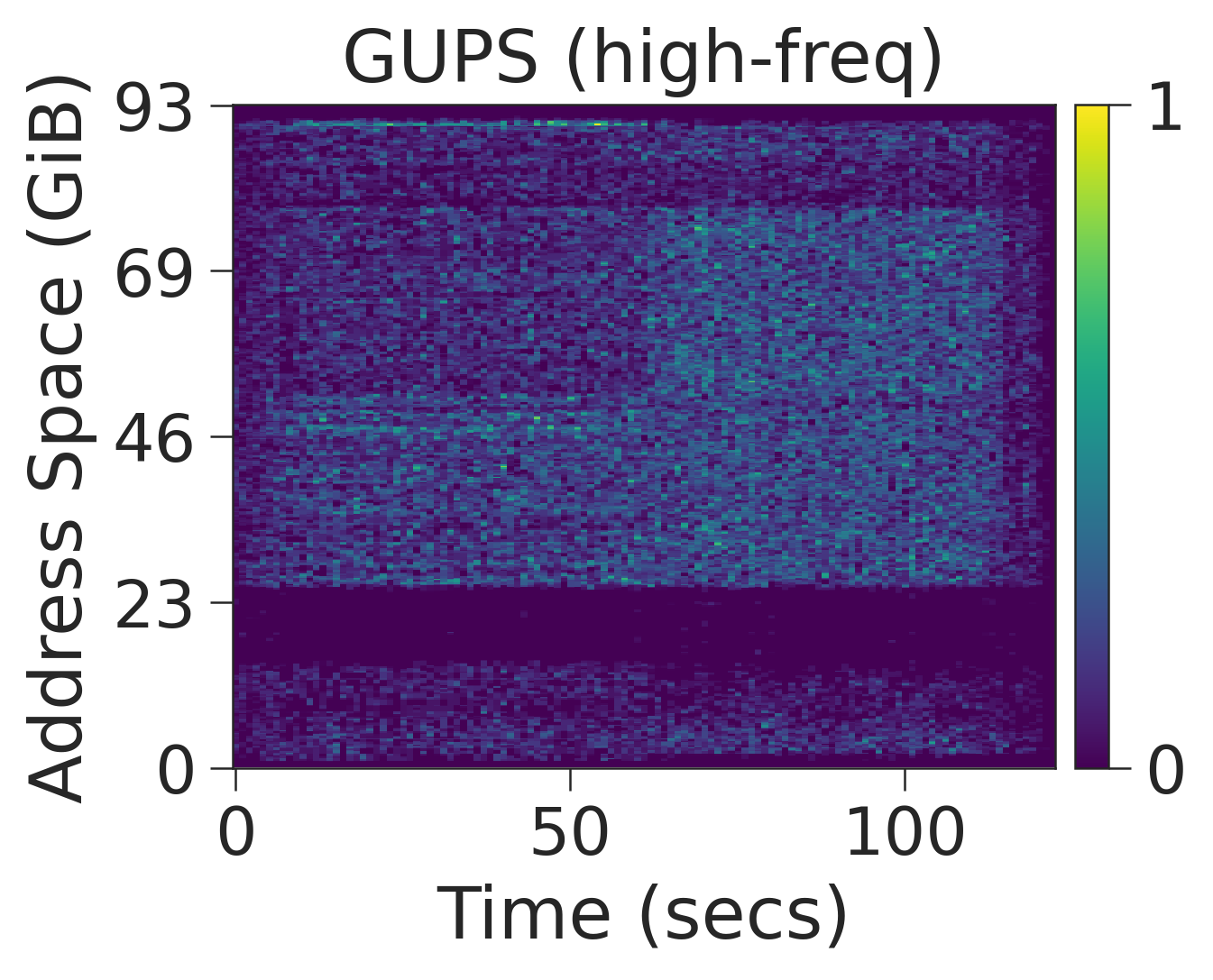}
    \end{minipage}
    \caption{Heatmap of GUPS generated by DAMON, using default HMSDK scanning frequency (left) and high scanning frequency (right)}
    \label{fig:gups-hmsdk}
    \vspace{-2ex}
\end{figure}

\smalltitle{GUPS:} We observe that it is not possible to improve HMSDK performance, which is surprising given that GUPS is a very simple workload.
After investigating further, we found that HMSDK is unable to differentiate between heavily accessed pages and lightly accessed pages irrespective of its configuration. This constitutes a major limitation of DAMON, which splits the address space into regions and assumes that all pages within a region have similar access frequency. Unfortunately, the hot pages of GUPS are distributed across the address space. Thus, DAMON finds all the regions to be hot. Figure~\ref{fig:gups-hmsdk} shows the heatmaps generated by DAMON for GUPS with two different sets of monitoring parameters. Irrespective of the parameter values, DAMON is unable to identify hot and cold pages accurately.

\vspace{-1ex}
\begin{mdframed}[linecolor=black,linewidth=1pt,innerleftmargin=8pt,innerrightmargin=8pt,innertopmargin=4pt,innerbottommargin=4pt, skipabove=8pt]
\textbf{Takeaway: }\textit{HMSDK experiences issues similar to HeMem, despite their apparent dissimilar design. HMSDK issue stem from static thresholds, making it impossible to reflect the application behavior.}
\end{mdframed}
\vspace{1ex}


\subsection{Comparison to Memtis}
\label{sec:memtis-eval}

Memtis~\cite{memtis2023} is a state-of-the-art memory tiering system that improves upon HeMem.
Memtis also employs PEBS to monitor page accesses and runs migration periodically in the background. Memtis' core improvements over HeMem are:
\begin{enumerate}[leftmargin=1.2em]
    \item It \textit{dynamically adapts hot thresholds} so as to maintain the hot set size close to the fast tier capacity.
    \item Introduces a \textit{warm} class for page classification. Memtis avoids migrating warm pages when the migration overhead would overshadow the benefit.
    \item Memtis dynamically chooses the page size (hugepage versus regular page) based on subpage access skewness. If Memtis finds that only a small subset of pages in a huge-page are hot, it splits the huge-pages into base pages and promotes only the hot base pages.
\end{enumerate}

\begin{figure}
    \centering
    \setlength{\abovecaptionskip}{0pt}
    \includegraphics[width=\linewidth]{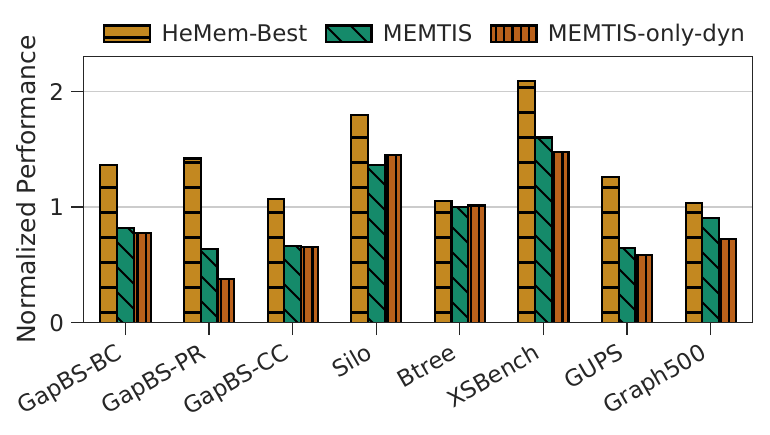}
    \caption{Performance of HeMem's best config found vs Memtis, normalized to HeMem default config performance.}
    \label{fig:memtis}
    \vspace{-3ex}
\end{figure}

We compare Memtis' performance with the one achieved by default and best HeMem configurations.
We include Memtis' performance, where we enable only the dynamically threshold adaptation feature and disable the warm pages and hugepage split features (i.e., MEMTIS-only-dyn). Figure~\ref{fig:memtis} shows the normalized performance for our workload set.

Similar to the results shown by authors of Memtis, we see that Memtis outperforms HeMem on default configuration for some workloads (e.g., Silo, XSBench), yet for most Memtis is slower.
More importantly, we observe that the best-performing HeMem configuration outperforms Memtis for all workloads.

Our analysis revealed multiple factors for the poor performance of Memtis.
We find that Memtis spends a significant amount of time in the kernel for page allocations, page splitting and migrations for these workloads. Second, Memtis uses very high sampling period (100K) for writes which leads to low write sampling accuracy. Therefore, Memtis performs poorly on workloads that are write-intensive. Third, Memtis uses static thresholds for other parameters such as cooling period, hot threshold adaptation period and migration period. By only dynamically adjusting the hot threshold, Memtis is unable to fully adapt to the workload. 

\begin{mdframed}[linecolor=black,linewidth=1pt,innerleftmargin=8pt,innerrightmargin=8pt,innertopmargin=4pt,innerbottommargin=4pt, skipabove=8pt]
\textbf{Takeaway: }\textit{Dynamically adjusting one or two knobs is beneficial but insufficient to extract maximum performance from tiering systems. For maximum performance, tiering systems need to adjust all parameters to reflect the application behavior.}
\end{mdframed}
\vspace{1ex}

\section{Discussion and future work}
\label{sec:discussion}

One of the key takeways from our work is that existing tiering solutions do not perform well under all circumstances. They fail to adapt to the workload and/or the underlying hardware. Going forward, we need to build tiering systems that can use high-level application knowledge and monitor hardware characteristics while making data placement decisions. Ideally, we should design memory tiering systems that are fully adaptable, i.e., have no knobs. However, designing such a complex system can be quite challenging. Below, we list some ways to get close to such a solution:

\textbf{Use programmer annotations/hints.} One of the main takeaways from our evaluation is that utilizing workload behavior to make data placement decisions in memory tiering systems can provide significant benefits. We realize this by using an optimizer which learns the application behavior and sets tiering parameters accordingly. However, using an optimizer has some limitations (e.g., it is time consuming, requires access to the application and inputs) and is not feasible to use in all scenarios. An easier way to provide high-level information to the tiering engine is through program annotation and hints. In some of the workloads we evaluated such as Graph500 and XSBench, we find that the hot and cold pages are from different allocations. Programmers can therefore provide hints at allocation time about the type of access: random or sequential and about the expected hotness which would help tiering systems make smart placement and migration decisions.

\textbf{Use cost-benefit models.} Prior work~\cite{cbmm2022} has argued that the kernel should use cost-benefit models to make policy decisions. We believe this argument holds for memory tiering systems as well. Through our evaluation, we find that existing systems~\cite{hemem2021,memtis2023,hmsdk2024} make sub-optimal policy decisions because they ignore the cost of page migrations. For many workloads (e.g. PageRank and XSBench), HeMem with the default configuration performs many high-cost non-beneficial page migrations. Migrations are expensive on machines with limited memory bandwidth or when applications are saturating the bandwidth. Page migrations, especially under memory pressure, has a detrimental impact on application performance~\cite{matryoshka2024}. Tiering systems should estimate the cost of page migration and benefit from the low latency access before moving pages between tiers. One way to estimate the cost is to measure the bandwidth utilization or memory queuing latencies using hardware performance counters.

\textbf{Identifying and adapting to application phases.} Many workloads have distinct execution phases. For instance, machine learning algorithms have a data loading phase, a preparation phase, and then in every iteration a forward propagation and back propagation. The memory access behavior is very different in each of these phases. Existing tiering systems are unable to identify and adapt to the different phases. For instance, with the Btree workload, we find that the optimizer fails to accelerate both the insert phase and the lookup phase. Since we set the parameter values at the beginning of the workload execution and do not change it dynamically during runtime, the tiering system does not perform optimally for both the phases.
Future tiering systems should identify and adapt to the program phases to achieve maximum performance.

\textbf{Intelligent tiering systems.} In our work, the optimizer was able to learn about the workload behavior and the appropriate parameter values by running multiple iterations of the workload. This could be replaced by a machine learning model that is trained on various workloads. In that case, the model could be trained to predict the right parameter values or even the exact placement and migration decisions for a given workload.
\section{Related Work}
In this section, we discuss some of the research in memory tiering relevant to our work.

\textbf{Page-based memory tiering systems:} Most of the previous proposed tiering systems perform page migrations transparently to the applications. These systems are based on heuristics and use static thresholds to make policy decisions. Some use recency-based policies~\cite{nimble2019,tpp2023} while others use frequency-based policies~\cite{thermostat2017,autotiering2017,hemem2021,tmts2023, hmsdk2024}. Some other use both page access recency and frequency information to make data placement decisions~\cite{tmts2023,multiclock2022,adaptivepagemigration2020}. We find that these systems are sub-optimal as they do not properly adapt to the workload or the underlying hardware.

\textbf{Dynamic tuning of knobs in tiering systems:} There has been some prior work that try to adapt parts of a tiering system to the workload. Memtis~\cite{memtis2023} uses dynamic threshold adaptation for page hotness classification which leads to better fast memory tier utilization. Cori~\cite{cori2021} tunes the periodicity of data movement in hybrid memory systems by extracting data reuse patterns from the application. Researches have proposed systems that migrate pages at different granularities for different workloads~\cite{granularityawaremigration2018,subpagemigration2021}. Our research finds that it is important to adapt all parts of a tiering system for maximum performance. 
Instead of adjusting the parameters, some tiering systems use different policies for different workloads. Heo et.al., propose a dynamic policy selection mechanism which identifies the best migration policy amongst LRU, LFU and random for a given workload~\cite{adaptivepagemigration2020}.
Yu et.al., build bandwidth-aware tiering systems~\cite{bwawaremigration}.

\textbf{Machine learning for data placement.} Sibyl~\cite{sibyl2022} uses reinforcement learning for data placement in hybrid storage systems. It observes different features of the running workload and storage devices such as number of accesses, access interval and free space to make data placement decisions. Kleio~\cite{kleio2019} uses deep neural networks to make intelligent page placement decision.

\textbf{Hardware tiering and profile-guided data placement.} To overcome the inaccuracies and inefficiencies in software profiling, Ramos et.al., propose augmenting the memory controller hardware to monitor access pattern and migrate pages between memories which would be more efficient~\cite{hwtiering2011}.

X-Mem~\cite{xmem2016} and Mira~\cite{mira2023} leverage profile guided techniques to determine object hotness offline and make data placement decisions during allocation time. These approaches do not work well for all applications, especially for those whose hot set changes over time.

\textbf{Optimizing knobs for systems.} Determining the right values for knobs has been a long-standing research problem in many systems. For example, LlamaTune \cite{kanellis2022llamatune} uses domain knowledge to efficiently sample and tune the knobs for DBMS with Bayesian Optimization. Cao et al. \cite{cao2018towards} study the black-box auto-tuning for storage systems and find that optimal configurations are highly dependent on hardware, software, and workloads; no single technique is superior to all others. Our work is the first to shed light on optimizing knobs for memory-tiering systems.

\section{Conclusion}

Existing tiering solutions that use heuristics and static thresholds do not perform well under all scenarios. The key insight from our work is that parameters (knobs) of existing tiering systems can be tweaked to express different data placement and migration patterns. Tuning parameter values can yield significant performance benefits for most workloads running on different types of hardware. We find that a Bayesian optimizer can tune the parameters effectively as it able to successfully generate a mental model of the program behavior. We observe that in almost all cases, the optimizer identifies the workload pattern correctly. By using an optimizer to tune the parameter knobs of existing tiering systems, we can achieve up to $2x$ performance improvement.

\balance

\bibliographystyle{ACM-Reference-Format}
\bibliography{references}


\begin{thebibliography}{50}


\ifx \showCODEN    \undefined \def \showCODEN     #1{\unskip}     \fi
\ifx \showDOI      \undefined \def \showDOI       #1{#1}\fi
\ifx \showISBNx    \undefined \def \showISBNx     #1{\unskip}     \fi
\ifx \showISBNxiii \undefined \def \showISBNxiii  #1{\unskip}     \fi
\ifx \showISSN     \undefined \def \showISSN      #1{\unskip}     \fi
\ifx \showLCCN     \undefined \def \showLCCN      #1{\unskip}     \fi
\ifx \shownote     \undefined \def \shownote      #1{#1}          \fi
\ifx \showarticletitle \undefined \def \showarticletitle #1{#1}   \fi
\ifx \showURL      \undefined \def \showURL       {\relax}        \fi
\providecommand\bibfield[2]{#2}
\providecommand\bibinfo[2]{#2}
\providecommand\natexlab[1]{#1}
\providecommand\showeprint[2][]{arXiv:#2}

\bibitem[Achermann and Panwar(2020)]%
        {reto2020btree}
\bibfield{author}{\bibinfo{person}{Reto Achermann} {and} \bibinfo{person}{Ashish Panwar}.} \bibinfo{year}{2020}\natexlab{}.
\newblock \showarticletitle{Mitosis Workload Btree}. \bibinfo{howpublished}{\url{https://github.com/mitosis-project/mitosis-workload-btree}}.
\newblock


\bibitem[Adavally and Kavi(2021)]%
        {subpagemigration2021}
\bibfield{author}{\bibinfo{person}{Shashank Adavally} {and} \bibinfo{person}{Krishna Kavi}.} \bibinfo{year}{2021}\natexlab{}.
\newblock \showarticletitle{Subpage migration in heterogeneous memory systems}. In \bibinfo{booktitle}{\emph{Workshop on Heterogeneous Memory Systems (HMEM-2021), Colocated with ICS 2021}}.
\newblock


\bibitem[Agarwal and Wenisch(2017)]%
        {thermostat2017}
\bibfield{author}{\bibinfo{person}{Neha Agarwal} {and} \bibinfo{person}{Thomas~F Wenisch}.} \bibinfo{year}{2017}\natexlab{}.
\newblock \showarticletitle{Thermostat: Application-transparent page management for two-tiered main memory}. In \bibinfo{booktitle}{\emph{Proceedings of the Twenty-Second International Conference on Architectural Support for Programming Languages and Operating Systems}}. \bibinfo{pages}{631--644}.
\newblock


\bibitem[Beamer et~al\mbox{.}(2015)]%
        {beamer2015gap}
\bibfield{author}{\bibinfo{person}{Scott Beamer}, \bibinfo{person}{Krste Asanovi{\'c}}, {and} \bibinfo{person}{David Patterson}.} \bibinfo{year}{2015}\natexlab{}.
\newblock \showarticletitle{The GAP benchmark suite}.
\newblock \bibinfo{journal}{\emph{arXiv preprint arXiv:1508.03619}} (\bibinfo{year}{2015}).
\newblock


\bibitem[Cao et~al\mbox{.}(2020)]%
        {carver2020}
\bibfield{author}{\bibinfo{person}{Zhen Cao}, \bibinfo{person}{Geoff Kuenning}, {and} \bibinfo{person}{Erez Zadok}.} \bibinfo{year}{2020}\natexlab{}.
\newblock \showarticletitle{Carver: Finding Important Parameters for Storage System Tuning}. In \bibinfo{booktitle}{\emph{18th USENIX Conference on File and Storage Technologies (FAST 20)}}. \bibinfo{publisher}{USENIX Association}, \bibinfo{address}{Santa Clara, CA}, \bibinfo{pages}{43--57}.
\newblock
\showISBNx{978-1-939133-12-0}
\urldef\tempurl%
\url{https://www.usenix.org/conference/fast20/presentation/cao-zhen}
\showURL{%
\tempurl}


\bibitem[Cao et~al\mbox{.}(2018)]%
        {cao2018towards}
\bibfield{author}{\bibinfo{person}{Zhen Cao}, \bibinfo{person}{Vasily Tarasov}, \bibinfo{person}{Sachin Tiwari}, {and} \bibinfo{person}{Erez Zadok}.} \bibinfo{year}{2018}\natexlab{}.
\newblock \showarticletitle{Towards Better Understanding of Black-box $\{$Auto-Tuning$\}$: A Comparative Analysis for Storage Systems}. In \bibinfo{booktitle}{\emph{2018 USENIX Annual Technical Conference (USENIX ATC 18)}}. \bibinfo{pages}{893--907}.
\newblock


\bibitem[Corporation.(2018)]%
        {pebs2018}
\bibfield{author}{\bibinfo{person}{Intel Corporation.}} \bibinfo{year}{2018}\natexlab{}.
\newblock \showarticletitle{Intel 64 and IA-32 Architectures Software Developer Manuals}.
\newblock \bibinfo{howpublished}{\url{https://software.intel.com/articles/intel-sdm}}.
\newblock  (\bibinfo{year}{2018}).
\newblock


\bibitem[Cutress and Tallis(2018)]%
        {nvm2018}
\bibfield{author}{\bibinfo{person}{Ian Cutress} {and} \bibinfo{person}{Billy Tallis}.} \bibinfo{year}{2018}\natexlab{}.
\newblock \showarticletitle{Intel Launches Optane DIMMs up to 512GB: Apache Pass is Here}. \bibinfo{howpublished}{\url{https://www.anandtech.com/show/12828/intel-launches-optane-dimms-up-to-512gb-apache-pass-is-here}}.
\newblock


\bibitem[Doudali et~al\mbox{.}(2019)]%
        {kleio2019}
\bibfield{author}{\bibinfo{person}{Thaleia~Dimitra Doudali}, \bibinfo{person}{Sergey Blagodurov}, \bibinfo{person}{Abhinav Vishnu}, \bibinfo{person}{Sudhanva Gurumurthi}, {and} \bibinfo{person}{Ada Gavrilovska}.} \bibinfo{year}{2019}\natexlab{}.
\newblock \showarticletitle{Kleio: A hybrid memory page scheduler with machine intelligence}. In \bibinfo{booktitle}{\emph{Proceedings of the 28th International Symposium on High-Performance Parallel and Distributed Computing}}. \bibinfo{pages}{37--48}.
\newblock


\bibitem[Doudali et~al\mbox{.}(2021)]%
        {cori2021}
\bibfield{author}{\bibinfo{person}{Thaleia~Dimitra Doudali}, \bibinfo{person}{Daniel Zahka}, {and} \bibinfo{person}{Ada Gavrilovska}.} \bibinfo{year}{2021}\natexlab{}.
\newblock \showarticletitle{Cori: Dancing to the right beat of periodic data movements over hybrid memory systems}. In \bibinfo{booktitle}{\emph{2021 IEEE International Parallel and Distributed Processing Symposium (IPDPS)}}. IEEE, \bibinfo{pages}{350--359}.
\newblock


\bibitem[Dulloor et~al\mbox{.}(2016)]%
        {xmem2016}
\bibfield{author}{\bibinfo{person}{Subramanya~R Dulloor}, \bibinfo{person}{Amitabha Roy}, \bibinfo{person}{Zheguang Zhao}, \bibinfo{person}{Narayanan Sundaram}, \bibinfo{person}{Nadathur Satish}, \bibinfo{person}{Rajesh Sankaran}, \bibinfo{person}{Jeff Jackson}, {and} \bibinfo{person}{Karsten Schwan}.} \bibinfo{year}{2016}\natexlab{}.
\newblock \showarticletitle{Data tiering in heterogeneous memory systems}. In \bibinfo{booktitle}{\emph{Proceedings of the Eleventh European Conference on Computer Systems}}. \bibinfo{pages}{1--16}.
\newblock


\bibitem[Duplyakin et~al\mbox{.}(2019)]%
        {cloudlab}
\bibfield{author}{\bibinfo{person}{Dmitry Duplyakin}, \bibinfo{person}{Robert Ricci}, \bibinfo{person}{Aleksander Maricq}, \bibinfo{person}{Gary Wong}, \bibinfo{person}{Jonathon Duerig}, \bibinfo{person}{Eric Eide}, \bibinfo{person}{Leigh Stoller}, \bibinfo{person}{Mike Hibler}, \bibinfo{person}{David Johnson}, \bibinfo{person}{Kirk Webb}, \bibinfo{person}{Aditya Akella}, \bibinfo{person}{Kuangching Wang}, \bibinfo{person}{Glenn Ricart}, \bibinfo{person}{Larry Landweber}, \bibinfo{person}{Chip Elliott}, \bibinfo{person}{Michael Zink}, \bibinfo{person}{Emmanuel Cecchet}, \bibinfo{person}{Snigdhaswin Kar}, {and} \bibinfo{person}{Prabodh Mishra}.} \bibinfo{year}{2019}\natexlab{}.
\newblock \showarticletitle{The Design and Operation of {CloudLab}}. In \bibinfo{booktitle}{\emph{2019 USENIX Annual Technical Conference (USENIX ATC 19)}}. \bibinfo{publisher}{USENIX Association}, \bibinfo{address}{Renton, WA}, \bibinfo{pages}{1--14}.
\newblock
\showISBNx{978-1-939133-03-8}
\urldef\tempurl%
\url{https://www.usenix.org/conference/atc19/presentation/duplyakin}
\showURL{%
\tempurl}


\bibitem[Duraisamy et~al\mbox{.}(2023)]%
        {tmts2023}
\bibfield{author}{\bibinfo{person}{Padmapriya Duraisamy}, \bibinfo{person}{Wei Xu}, \bibinfo{person}{Scott Hare}, \bibinfo{person}{Ravi Rajwar}, \bibinfo{person}{David Culler}, \bibinfo{person}{Zhiyi Xu}, \bibinfo{person}{Jianing Fan}, \bibinfo{person}{Christopher Kennelly}, \bibinfo{person}{Bill McCloskey}, \bibinfo{person}{Danijela Mijailovic}, {et~al\mbox{.}}} \bibinfo{year}{2023}\natexlab{}.
\newblock \showarticletitle{Towards an adaptable systems architecture for memory tiering at warehouse-scale}. In \bibinfo{booktitle}{\emph{Proceedings of the 28th ACM International Conference on Architectural Support for Programming Languages and Operating Systems, Volume 3}}. \bibinfo{pages}{727--741}.
\newblock


\bibitem[Fekry et~al\mbox{.}(2020)]%
        {tuneful2020}
\bibfield{author}{\bibinfo{person}{Ayat Fekry}, \bibinfo{person}{Lucian Carata}, \bibinfo{person}{Thomas Pasquier}, \bibinfo{person}{Andrew Rice}, {and} \bibinfo{person}{Andy Hopper}.} \bibinfo{year}{2020}\natexlab{}.
\newblock \showarticletitle{To Tune or Not to Tune? In Search of Optimal Configurations for Data Analytics}. In \bibinfo{booktitle}{\emph{Proceedings of the 26th ACM SIGKDD International Conference on Knowledge Discovery \& Data Mining}} (Virtual Event, CA, USA) \emph{(\bibinfo{series}{KDD '20})}. \bibinfo{publisher}{Association for Computing Machinery}, \bibinfo{address}{New York, NY, USA}, \bibinfo{pages}{2494–2504}.
\newblock
\showISBNx{9781450379984}
\urldef\tempurl%
\url{https://doi.org/10.1145/3394486.3403299}
\showDOI{\tempurl}


\bibitem[Freischuetz et~al\mbox{.}(2025)]%
        {tuna2025}
\bibfield{author}{\bibinfo{person}{Johannes Freischuetz}, \bibinfo{person}{Konstantinos Kanellis}, \bibinfo{person}{Brian Kroth}, {and} \bibinfo{person}{Shivaram Venkataraman}.} \bibinfo{year}{2025}\natexlab{}.
\newblock \showarticletitle{TUNA: Tuning Unstable and Noisy Cloud Applications}. In \bibinfo{booktitle}{\emph{Proceedings of the Twentieth European Conference on Computer Systems}} (Rotterdam, Netherlands) \emph{(\bibinfo{series}{EuroSys '25})}. \bibinfo{publisher}{Association for Computing Machinery}, \bibinfo{address}{New York, NY, USA}, \bibinfo{pages}{954–973}.
\newblock
\showISBNx{9798400711961}
\urldef\tempurl%
\url{https://doi.org/10.1145/3689031.3717480}
\showDOI{\tempurl}


\bibitem[Guo et~al\mbox{.}(2023)]%
        {mira2023}
\bibfield{author}{\bibinfo{person}{Zhiyuan Guo}, \bibinfo{person}{Zijian He}, {and} \bibinfo{person}{Yiying Zhang}.} \bibinfo{year}{2023}\natexlab{}.
\newblock \showarticletitle{Mira: A program-behavior-guided far memory system}. In \bibinfo{booktitle}{\emph{Proceedings of the 29th Symposium on Operating Systems Principles}}. \bibinfo{pages}{692--708}.
\newblock


\bibitem[Heo et~al\mbox{.}(2020)]%
        {adaptivepagemigration2020}
\bibfield{author}{\bibinfo{person}{Taekyung Heo}, \bibinfo{person}{Yang Wang}, \bibinfo{person}{Wei Cui}, \bibinfo{person}{Jaehyuk Huh}, {and} \bibinfo{person}{Lintao Zhang}.} \bibinfo{year}{2020}\natexlab{}.
\newblock \showarticletitle{Adaptive page migration policy with huge pages in tiered memory systems}.
\newblock \bibinfo{journal}{\emph{IEEE Trans. Comput.}} \bibinfo{volume}{71}, \bibinfo{number}{1} (\bibinfo{year}{2020}), \bibinfo{pages}{53--68}.
\newblock


\bibitem[Hutter et~al\mbox{.}(2011)]%
        {hutter2011sequential}
\bibfield{author}{\bibinfo{person}{Frank Hutter}, \bibinfo{person}{Holger~H Hoos}, {and} \bibinfo{person}{Kevin Leyton-Brown}.} \bibinfo{year}{2011}\natexlab{}.
\newblock \showarticletitle{Sequential model-based optimization for general algorithm configuration}. In \bibinfo{booktitle}{\emph{Learning and Intelligent Optimization: 5th International Conference, LION 5, Rome, Italy, January 17-21, 2011. Selected Papers 5}}. Springer, \bibinfo{pages}{507--523}.
\newblock


\bibitem[Inc.(2020)]%
        {cxlspec}
\bibfield{author}{\bibinfo{person}{Compute Express Link~Consortium. Inc.}} \bibinfo{year}{2020}\natexlab{}.
\newblock \showarticletitle{CXL® Specification}. \bibinfo{howpublished}{\url{https://computeexpresslink.org/cxl-specification/}}.
\newblock


\bibitem[Izraelevitz et~al\mbox{.}(2019)]%
        {nvmperf2019}
\bibfield{author}{\bibinfo{person}{Joseph Izraelevitz}, \bibinfo{person}{Jian Yang}, \bibinfo{person}{Lu Zhang}, \bibinfo{person}{Juno Kim}, \bibinfo{person}{Xiao Liu}, \bibinfo{person}{Amirsaman Memaripour}, \bibinfo{person}{Yun~Joon Soh}, \bibinfo{person}{Zixuan Wang}, \bibinfo{person}{Yi Xu}, \bibinfo{person}{Subramanya~R Dulloor}, {et~al\mbox{.}}} \bibinfo{year}{2019}\natexlab{}.
\newblock \showarticletitle{Basic performance measurements of the intel optane DC persistent memory module}.
\newblock \bibinfo{journal}{\emph{arXiv preprint arXiv:1903.05714}} (\bibinfo{year}{2019}).
\newblock


\bibitem[Kanellis et~al\mbox{.}(2020)]%
        {kanellis2020too}
\bibfield{author}{\bibinfo{person}{Konstantinos Kanellis}, \bibinfo{person}{Ramnatthan Alagappan}, {and} \bibinfo{person}{Shivaram Venkataraman}.} \bibinfo{year}{2020}\natexlab{}.
\newblock \showarticletitle{Too many knobs to tune? towards faster database tuning by pre-selecting important knobs}. In \bibinfo{booktitle}{\emph{12th USENIX Workshop on Hot Topics in Storage and File Systems (HotStorage 20)}}.
\newblock


\bibitem[Kanellis et~al\mbox{.}(2022)]%
        {kanellis2022llamatune}
\bibfield{author}{\bibinfo{person}{Konstantinos Kanellis}, \bibinfo{person}{Cong Ding}, \bibinfo{person}{Brian Kroth}, \bibinfo{person}{Andreas M{\"u}ller}, \bibinfo{person}{Carlo Curino}, {and} \bibinfo{person}{Shivaram Venkataraman}.} \bibinfo{year}{2022}\natexlab{}.
\newblock \showarticletitle{LlamaTune: Sample-efficient DBMS configuration tuning}.
\newblock \bibinfo{journal}{\emph{arXiv preprint arXiv:2203.05128}} (\bibinfo{year}{2022}).
\newblock


\bibitem[Lee et~al\mbox{.}(2024)]%
        {hmsdk2024}
\bibfield{author}{\bibinfo{person}{KyungSoo Lee}, \bibinfo{person}{Sohyun Kim}, \bibinfo{person}{Joohee Lee}, \bibinfo{person}{Donguk Moon}, \bibinfo{person}{Rakie Kim}, \bibinfo{person}{Honggyu Kim}, \bibinfo{person}{Hyeongtak Ji}, \bibinfo{person}{Yunjeong Mun}, {and} \bibinfo{person}{Youngpyo Joo}.} \bibinfo{year}{2024}\natexlab{}.
\newblock \showarticletitle{Improving key-value cache performance with heterogeneous memory tiering: A case study of CXL-based memory expansion}.
\newblock \bibinfo{journal}{\emph{IEEE Micro}} (\bibinfo{year}{2024}).
\newblock


\bibitem[Lee et~al\mbox{.}(2023)]%
        {memtis2023}
\bibfield{author}{\bibinfo{person}{Taehyung Lee}, \bibinfo{person}{Sumit~Kumar Monga}, \bibinfo{person}{Changwoo Min}, {and} \bibinfo{person}{Young~Ik Eom}.} \bibinfo{year}{2023}\natexlab{}.
\newblock \showarticletitle{MEMTIS: Efficient Memory Tiering with Dynamic Page Classification and Page Size Determination}. In \bibinfo{booktitle}{\emph{Proceedings of the 29th Symposium on Operating Systems Principles}}. \bibinfo{pages}{17--34}.
\newblock


\bibitem[Lepers and Zwaenepoel(2023)]%
        {johnnyCache2023}
\bibfield{author}{\bibinfo{person}{Baptiste Lepers} {and} \bibinfo{person}{Willy Zwaenepoel}.} \bibinfo{year}{2023}\natexlab{}.
\newblock \showarticletitle{Johnny Cache: the End of $\{$DRAM$\}$ Cache Conflicts (in Tiered Main Memory Systems)}. In \bibinfo{booktitle}{\emph{17th USENIX Symposium on Operating Systems Design and Implementation (OSDI 23)}}. \bibinfo{pages}{519--534}.
\newblock


\bibitem[Mansi et~al\mbox{.}(2022)]%
        {cbmm2022}
\bibfield{author}{\bibinfo{person}{Mark Mansi}, \bibinfo{person}{Bijan Tabatabai}, {and} \bibinfo{person}{Michael~M Swift}.} \bibinfo{year}{2022}\natexlab{}.
\newblock \showarticletitle{$\{$CBMM$\}$: Financial Advice for Kernel Memory Managers}. In \bibinfo{booktitle}{\emph{2022 USENIX Annual Technical Conference (USENIX ATC 22)}}. \bibinfo{pages}{593--608}.
\newblock


\bibitem[Maruf et~al\mbox{.}(2022)]%
        {multiclock2022}
\bibfield{author}{\bibinfo{person}{Adnan Maruf}, \bibinfo{person}{Ashikee Ghosh}, \bibinfo{person}{Janki Bhimani}, \bibinfo{person}{Daniel Campello}, \bibinfo{person}{Andy Rudoff}, {and} \bibinfo{person}{Raju Rangaswami}.} \bibinfo{year}{2022}\natexlab{}.
\newblock \showarticletitle{MULTI-CLOCK: Dynamic Tiering for Hybrid Memory Systems.}. In \bibinfo{booktitle}{\emph{HPCA}}. \bibinfo{pages}{925--937}.
\newblock


\bibitem[Maruf et~al\mbox{.}(2023)]%
        {tpp2023}
\bibfield{author}{\bibinfo{person}{Hasan~Al Maruf}, \bibinfo{person}{Hao Wang}, \bibinfo{person}{Abhishek Dhanotia}, \bibinfo{person}{Johannes Weiner}, \bibinfo{person}{Niket Agarwal}, \bibinfo{person}{Pallab Bhattacharya}, \bibinfo{person}{Chris Petersen}, \bibinfo{person}{Mosharaf Chowdhury}, \bibinfo{person}{Shobhit Kanaujia}, {and} \bibinfo{person}{Prakash Chauhan}.} \bibinfo{year}{2023}\natexlab{}.
\newblock \showarticletitle{TPP: Transparent page placement for CXL-enabled tiered-memory}. In \bibinfo{booktitle}{\emph{Proceedings of the 28th ACM International Conference on Architectural Support for Programming Languages and Operating Systems, Volume 3}}. \bibinfo{pages}{742--755}.
\newblock


\bibitem[Mo et~al\mbox{.}(2023)]%
        {kv-rl-tuning23}
\bibfield{author}{\bibinfo{person}{Dingheng Mo}, \bibinfo{person}{Fanchao Chen}, \bibinfo{person}{Siqiang Luo}, {and} \bibinfo{person}{Caihua Shan}.} \bibinfo{year}{2023}\natexlab{}.
\newblock \showarticletitle{Learning to Optimize LSM-trees: Towards A Reinforcement Learning based Key-Value Store for Dynamic Workloads}.
\newblock \bibinfo{journal}{\emph{Proc. ACM Manag. Data}} \bibinfo{volume}{1}, \bibinfo{number}{3}, Article \bibinfo{articleno}{213} (\bibinfo{date}{Nov.} \bibinfo{year}{2023}), \bibinfo{numpages}{25}~pages.
\newblock
\urldef\tempurl%
\url{https://doi.org/10.1145/3617333}
\showDOI{\tempurl}


\bibitem[Morgan(2020)]%
        {azurememorycosts}
\bibfield{author}{\bibinfo{person}{Timothy Morgan}.} \bibinfo{year}{2020}\natexlab{}.
\newblock \showarticletitle{CXL And Gen-Z Iron Out A Coherent Interconnect Strategy}. \bibinfo{howpublished}{\url{https://www.nextplatform.com/2020/04/03/cxl-and-gen-z-iron-out-a-coherent-interconnect-strategy/}}. In \bibinfo{booktitle}{\emph{The Next Platform}}.
\newblock


\bibitem[Murphy et~al\mbox{.}(2010)]%
        {murphy2010introducing}
\bibfield{author}{\bibinfo{person}{Richard~C Murphy}, \bibinfo{person}{Kyle~B Wheeler}, \bibinfo{person}{Brian~W Barrett}, {and} \bibinfo{person}{James~A Ang}.} \bibinfo{year}{2010}\natexlab{}.
\newblock \showarticletitle{Introducing the graph 500}.
\newblock \bibinfo{journal}{\emph{Cray Users Group (CUG)}} \bibinfo{volume}{19}, \bibinfo{number}{45-74} (\bibinfo{year}{2010}), \bibinfo{pages}{22}.
\newblock


\bibitem[Park et~al\mbox{.}(2019)]%
        {damon2019}
\bibfield{author}{\bibinfo{person}{SeongJae Park}, \bibinfo{person}{Yunjae Lee}, {and} \bibinfo{person}{Heon~Y Yeom}.} \bibinfo{year}{2019}\natexlab{}.
\newblock \showarticletitle{Profiling dynamic data access patterns with controlled overhead and quality}. In \bibinfo{booktitle}{\emph{Proceedings of the 20th International Middleware Conference Industrial Track}}. \bibinfo{pages}{1--7}.
\newblock


\bibitem[Plimpton et~al\mbox{.}(2006)]%
        {plimpton2006simple}
\bibfield{author}{\bibinfo{person}{Steven~J Plimpton}, \bibinfo{person}{Ron Brightwell}, \bibinfo{person}{Courtenay Vaughan}, \bibinfo{person}{Keith Underwood}, {and} \bibinfo{person}{Mike Davis}.} \bibinfo{year}{2006}\natexlab{}.
\newblock \showarticletitle{A simple synchronous distributed-memory algorithm for the HPCC RandomAccess benchmark}. In \bibinfo{booktitle}{\emph{2006 IEEE International Conference on Cluster Computing}}. IEEE, \bibinfo{pages}{1--7}.
\newblock


\bibitem[Ramos et~al\mbox{.}(2011)]%
        {hwtiering2011}
\bibfield{author}{\bibinfo{person}{Luiz~E Ramos}, \bibinfo{person}{Eugene Gorbatov}, {and} \bibinfo{person}{Ricardo Bianchini}.} \bibinfo{year}{2011}\natexlab{}.
\newblock \showarticletitle{Page placement in hybrid memory systems}. In \bibinfo{booktitle}{\emph{Proceedings of the international conference on Supercomputing}}. \bibinfo{pages}{85--95}.
\newblock


\bibitem[Raybuck et~al\mbox{.}(2021)]%
        {hemem2021}
\bibfield{author}{\bibinfo{person}{Amanda Raybuck}, \bibinfo{person}{Tim Stamler}, \bibinfo{person}{Wei Zhang}, \bibinfo{person}{Mattan Erez}, {and} \bibinfo{person}{Simon Peter}.} \bibinfo{year}{2021}\natexlab{}.
\newblock \showarticletitle{Hemem: Scalable tiered memory management for big data applications and real nvm}. In \bibinfo{booktitle}{\emph{Proceedings of the ACM SIGOPS 28th Symposium on Operating Systems Principles}}. \bibinfo{pages}{392--407}.
\newblock


\bibitem[Ryoo et~al\mbox{.}(2018)]%
        {granularityawaremigration2018}
\bibfield{author}{\bibinfo{person}{Jee~Ho Ryoo}, \bibinfo{person}{Lizy~K John}, {and} \bibinfo{person}{Arkaprava Basu}.} \bibinfo{year}{2018}\natexlab{}.
\newblock \showarticletitle{A case for granularity aware page migration}. In \bibinfo{booktitle}{\emph{Proceedings of the 2018 International Conference on Supercomputing}}. \bibinfo{pages}{352--362}.
\newblock


\bibitem[Shahriari et~al\mbox{.}(2015)]%
        {bo-survey}
\bibfield{author}{\bibinfo{person}{Bobak Shahriari}, \bibinfo{person}{Kevin Swersky}, \bibinfo{person}{Ziyu Wang}, \bibinfo{person}{Ryan~P Adams}, {and} \bibinfo{person}{Nando De~Freitas}.} \bibinfo{year}{2015}\natexlab{}.
\newblock \showarticletitle{Taking the human out of the loop: A review of Bayesian optimization}.
\newblock \bibinfo{journal}{\emph{Proc. IEEE}} \bibinfo{volume}{104}, \bibinfo{number}{1} (\bibinfo{year}{2015}), \bibinfo{pages}{148--175}.
\newblock


\bibitem[Singh et~al\mbox{.}(2022)]%
        {sibyl2022}
\bibfield{author}{\bibinfo{person}{Gagandeep Singh}, \bibinfo{person}{Rakesh Nadig}, \bibinfo{person}{Jisung Park}, \bibinfo{person}{Rahul Bera}, \bibinfo{person}{Nastaran Hajinazar}, \bibinfo{person}{David Novo}, \bibinfo{person}{Juan G{\'o}mez-Luna}, \bibinfo{person}{Sander Stuijk}, \bibinfo{person}{Henk Corporaal}, {and} \bibinfo{person}{Onur Mutlu}.} \bibinfo{year}{2022}\natexlab{}.
\newblock \showarticletitle{Sibyl: Adaptive and extensible data placement in hybrid storage systems using online reinforcement learning}. In \bibinfo{booktitle}{\emph{Proceedings of the 49th Annual International Symposium on Computer Architecture}}. \bibinfo{pages}{320--336}.
\newblock


\bibitem[Snoek et~al\mbox{.}(2012)]%
        {snoek2012practical}
\bibfield{author}{\bibinfo{person}{Jasper Snoek}, \bibinfo{person}{Hugo Larochelle}, {and} \bibinfo{person}{Ryan~P Adams}.} \bibinfo{year}{2012}\natexlab{}.
\newblock \showarticletitle{Practical bayesian optimization of machine learning algorithms}.
\newblock \bibinfo{journal}{\emph{Advances in Neural Information Processing Systems}}  \bibinfo{volume}{25} (\bibinfo{year}{2012}).
\newblock


\bibitem[Sun et~al\mbox{.}(2023)]%
        {cxlperf2023}
\bibfield{author}{\bibinfo{person}{Yan Sun}, \bibinfo{person}{Yifan Yuan}, \bibinfo{person}{Zeduo Yu}, \bibinfo{person}{Reese Kuper}, \bibinfo{person}{Chihun Song}, \bibinfo{person}{Jinghan Huang}, \bibinfo{person}{Houxiang Ji}, \bibinfo{person}{Siddharth Agarwal}, \bibinfo{person}{Jiaqi Lou}, \bibinfo{person}{Ipoom Jeong}, {et~al\mbox{.}}} \bibinfo{year}{2023}\natexlab{}.
\newblock \showarticletitle{Demystifying cxl memory with genuine cxl-ready systems and devices}. In \bibinfo{booktitle}{\emph{Proceedings of the 56th Annual IEEE/ACM International Symposium on Microarchitecture}}. \bibinfo{pages}{105--121}.
\newblock


\bibitem[Tramm et~al\mbox{.}(2014)]%
        {tramm2014xsbench}
\bibfield{author}{\bibinfo{person}{John~R Tramm}, \bibinfo{person}{Andrew~R Siegel}, \bibinfo{person}{Tanzima Islam}, {and} \bibinfo{person}{Martin Schulz}.} \bibinfo{year}{2014}\natexlab{}.
\newblock \showarticletitle{XSBench-the development and verification of a performance abstraction for Monte Carlo reactor analysis}.
\newblock \bibinfo{journal}{\emph{The Role of Reactor Physics toward a Sustainable Future (PHYSOR)}} (\bibinfo{year}{2014}).
\newblock


\bibitem[Tu et~al\mbox{.}(2013)]%
        {tu2013speedy}
\bibfield{author}{\bibinfo{person}{Stephen Tu}, \bibinfo{person}{Wenting Zheng}, \bibinfo{person}{Eddie Kohler}, \bibinfo{person}{Barbara Liskov}, {and} \bibinfo{person}{Samuel Madden}.} \bibinfo{year}{2013}\natexlab{}.
\newblock \showarticletitle{Speedy transactions in multicore in-memory databases}. In \bibinfo{booktitle}{\emph{Proceedings of the Twenty-Fourth ACM Symposium on Operating Systems Principles}}. \bibinfo{pages}{18--32}.
\newblock


\bibitem[Weiner et~al\mbox{.}(2022)]%
        {tmo2022}
\bibfield{author}{\bibinfo{person}{Johannes Weiner}, \bibinfo{person}{Niket Agarwal}, \bibinfo{person}{Dan Schatzberg}, \bibinfo{person}{Leon Yang}, \bibinfo{person}{Hao Wang}, \bibinfo{person}{Blaise Sanouillet}, \bibinfo{person}{Bikash Sharma}, \bibinfo{person}{Tejun Heo}, \bibinfo{person}{Mayank Jain}, \bibinfo{person}{Chunqiang Tang}, {et~al\mbox{.}}} \bibinfo{year}{2022}\natexlab{}.
\newblock \showarticletitle{TMO: Transparent memory offloading in datacenters}. In \bibinfo{booktitle}{\emph{Proceedings of the 27th ACM International Conference on Architectural Support for Programming Languages and Operating Systems}}. \bibinfo{pages}{609--621}.
\newblock


\bibitem[Xiang et~al\mbox{.}(2024)]%
        {matryoshka2024}
\bibfield{author}{\bibinfo{person}{Lingfeng Xiang}, \bibinfo{person}{Zhen Lin}, \bibinfo{person}{Weishu Deng}, \bibinfo{person}{Hui Lu}, \bibinfo{person}{Jia Rao}, \bibinfo{person}{Yifan Yuan}, {and} \bibinfo{person}{Ren Wang}.} \bibinfo{year}{2024}\natexlab{}.
\newblock \showarticletitle{MATRYOSHKA: Non-Exclusive Memory Tiering via Transactional Page Migration}.
\newblock \bibinfo{journal}{\emph{arXiv preprint arXiv:2401.13154}} (\bibinfo{year}{2024}).
\newblock


\bibitem[Xu et~al\mbox{.}(2015)]%
        {xu2015performance}
\bibfield{author}{\bibinfo{person}{Qiumin Xu}, \bibinfo{person}{Huzefa Siyamwala}, \bibinfo{person}{Mrinmoy Ghosh}, \bibinfo{person}{Tameesh Suri}, \bibinfo{person}{Manu Awasthi}, \bibinfo{person}{Zvika Guz}, \bibinfo{person}{Anahita Shayesteh}, {and} \bibinfo{person}{Vijay Balakrishnan}.} \bibinfo{year}{2015}\natexlab{}.
\newblock \showarticletitle{Performance analysis of NVMe SSDs and their implication on real world databases}. In \bibinfo{booktitle}{\emph{Proceedings of the 8th ACM International Systems and Storage Conference}}. \bibinfo{pages}{1--11}.
\newblock


\bibitem[Yan et~al\mbox{.}(2019)]%
        {nimble2019}
\bibfield{author}{\bibinfo{person}{Zi Yan}, \bibinfo{person}{Daniel Lustig}, \bibinfo{person}{David Nellans}, {and} \bibinfo{person}{Abhishek Bhattacharjee}.} \bibinfo{year}{2019}\natexlab{}.
\newblock \showarticletitle{Nimble page management for tiered memory systems}. In \bibinfo{booktitle}{\emph{Proceedings of the Twenty-Fourth International Conference on Architectural Support for Programming Languages and Operating Systems}}. \bibinfo{pages}{331--345}.
\newblock


\bibitem[Yang et~al\mbox{.}(2017)]%
        {autotiering2017}
\bibfield{author}{\bibinfo{person}{Zhengyu Yang}, \bibinfo{person}{Morteza Hoseinzadeh}, \bibinfo{person}{Allen Andrews}, \bibinfo{person}{Clay Mayers}, \bibinfo{person}{David~Thomas Evans}, \bibinfo{person}{Rory~Thomas Bolt}, \bibinfo{person}{Janki Bhimani}, \bibinfo{person}{Ningfang Mi}, {and} \bibinfo{person}{Steven Swanson}.} \bibinfo{year}{2017}\natexlab{}.
\newblock \showarticletitle{AutoTiering: Automatic data placement manager in multi-tier all-flash datacenter}. In \bibinfo{booktitle}{\emph{2017 IEEE 36th International Performance Computing and Communications Conference (IPCCC)}}. IEEE, \bibinfo{pages}{1--8}.
\newblock


\bibitem[Yu et~al\mbox{.}(2017)]%
        {bwawaremigration}
\bibfield{author}{\bibinfo{person}{Seongdae Yu}, \bibinfo{person}{Seongbeom Park}, {and} \bibinfo{person}{Woongki Baek}.} \bibinfo{year}{2017}\natexlab{}.
\newblock \showarticletitle{Design and implementation of bandwidth-aware memory placement and migration policies for heterogeneous memory systems}. In \bibinfo{booktitle}{\emph{Proceedings of the International Conference on Supercomputing}}. \bibinfo{pages}{1--10}.
\newblock


\bibitem[Zhang et~al\mbox{.}(2020)]%
        {optimal2020}
\bibfield{author}{\bibinfo{person}{Lei Zhang}, \bibinfo{person}{Reza Karimi}, \bibinfo{person}{Irfan Ahmad}, {and} \bibinfo{person}{Ymir Vigfusson}.} \bibinfo{year}{2020}\natexlab{}.
\newblock \showarticletitle{Optimal data placement for heterogeneous cache, memory, and storage systems}.
\newblock \bibinfo{journal}{\emph{Proceedings of the ACM on Measurement and Analysis of Computing Systems}} \bibinfo{volume}{4}, \bibinfo{number}{1} (\bibinfo{year}{2020}), \bibinfo{pages}{1--27}.
\newblock


\bibitem[Zhao et~al\mbox{.}(2022)]%
        {dremel22}
\bibfield{author}{\bibinfo{person}{Chenxingyu Zhao}, \bibinfo{person}{Tapan Chugh}, \bibinfo{person}{Jaehong Min}, \bibinfo{person}{Ming Liu}, {and} \bibinfo{person}{Arvind Krishnamurthy}.} \bibinfo{year}{2022}\natexlab{}.
\newblock \showarticletitle{Dremel: Adaptive Configuration Tuning of RocksDB KV-Store}.
\newblock \bibinfo{journal}{\emph{Proc. ACM Meas. Anal. Comput. Syst.}} \bibinfo{volume}{6}, \bibinfo{number}{2}, Article \bibinfo{articleno}{37} (\bibinfo{date}{June} \bibinfo{year}{2022}), \bibinfo{numpages}{30}~pages.
\newblock
\urldef\tempurl%
\url{https://doi.org/10.1145/3530903}
\showDOI{\tempurl}


\end{thebibliography}

\end{document}